# Spatial and Ecological Scaling of Stability in Spatial Community Networks

**Javier Jarillo[1]\*[†], Francisco J. Cao-García[2,3], Frederik De Laender[1]**

[1]Research Unit of Environmental and Evolutionary Biology, Namur Institute of Complex Systems, and Institute of Life, Earth, and the Environment, University of Namur, Rue de Bruxelles 61, Namur, 5000, Belgium.

[2]Departamento de Estructura de la Materia, Física Térmica y Electrónica, Universidad Complutense de Madrid, Plaza de Ciencias, 1, 28040 Madrid, Spain.

[3]Instituto Madrileño de Estudios Avanzados en Nanociencia, IMDEA Nanociencia, Calle Faraday, 9, 28049 Madrid, Spain.

**\* Correspondence:**
Corresponding Author
jjarillo@ucm.es

**† Present Address**:
Departamento de Estadística e Investigación Operativa, Universidad Complutense de Madrid, Plaza de Ciencias, 3, 28040 Madrid, Spain

## Abstract

There are many scales at which to quantify stability in spatial and ecological networks. Local-scale analyses focus on specific nodes of the spatial network, while regional-scale analyses consider the whole network. Similarly, species- and community-level analyses either account for single species or for the whole community. Furthermore, stability itself can be defined in multiple ways, including resistance (the inverse of the relative displacement caused by a perturbation), initial resilience (the rate of return after a perturbation), and invariability (the inverse of the relative amplitude of the population fluctuations). Here, we analyze the scale-dependence of these stability properties. More specifically, we ask how spatial scale (local vs regional) and ecological scale (species vs community) influence these stability properties. We find that regional initial resilience is the weighted arithmetic mean of the local initial resiliences. The regional resistance is the harmonic mean of local resistances, which makes regional resistance particularly vulnerable to nodes with low stability, unlike regional initial resilience. Analogous results hold for the relationship between community- and species-level initial resilience and resistance. Both resistance and initial resilience are "scale-free" properties: regional and community values are simply the biomass-weighted means of the local and species values, respectively. Thus, one can easily estimate both stability metrics of whole networks from partial sampling. In contrast, invariability generally is greater at the regional and community-level than at the local and species-level, respectively. Hence, estimating the invariability of spatial or ecological networks from measurements at the local or species level is more complicated, requiring an unbiased estimate of the network (*i.e.* region or community) size. In conclusion, we find that scaling of stability depends on the metric considered, and we present a reliable framework to estimate these metrics.

## 1 Introduction

Ecological stability is a property that can be broadly defined as the ability of an ecosystem to remain unaltered when challenged by perturbations. However, there exist multiple ways of characterizing stability, which leads to different stability definitions or components (Pimm, 1984; Grimm and Wissel, 1997; McCann, 2000). Different components include resistance (to perturbation), initial resilience (*i.e.*, the ability to recover from a perturbation) or invariability (*i.e.*, the ability to remain unaltered to repeated perturbations) (Fig. 1A). Different stability components can also vary with scale, impeding cross-system comparison of stability, or be scale-independent instead (Levin, 1992; Wang et al., 2017; Domínguez-García et al., 2019; Kéfi et al., 2019; Greig et al., 2022) (Fig. 1B). Thus, the ecological and spatial scale at which one studies an ecological system can be hypothesized to influence stability assessments.

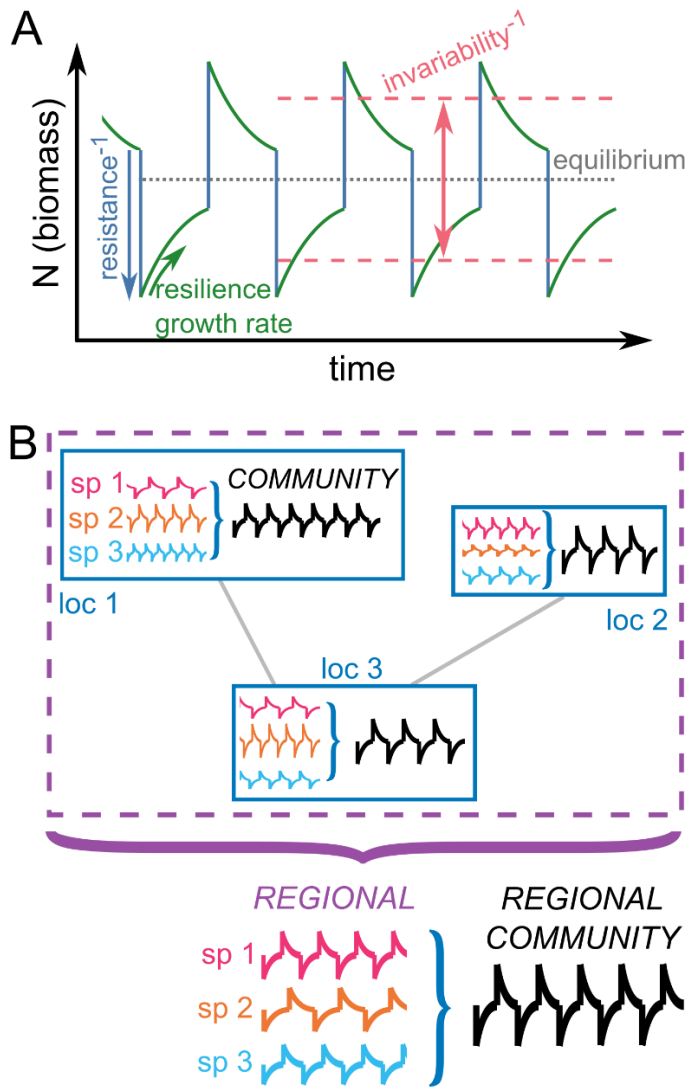

**Figure 1: (A)** Different stability components considered throughout the manuscript. In solid lines, we represent the hypothetical dynamics of the biomass of one species affected by multiple periodic disturbances. Resistance (blue lines) is defined as the inverse of the change of biomass as a consequence of the direct effect of the disturbance. The growth rate and the initial resilience of the system (green lines) both quantify the rate at which the system tends to recover to the equilibrium after the effect of the disturbance. The main difference between growth rate and initial resilience is whether it is computed relatively to the biomass prior the disturbance (growth rate), or relatively to the distance between the biomass after the perturbation and the equilibrium biomass (initial resilience). In both cases, we will consider the short-term response, in contrast to the asymptotic resilience (not considered in this manuscript). Finally, invariability (red dashed lines) represents the ability of the systems to remain close to the equilibrium when multiple disturbances affect (periodically or randomly) the species. Panel inspired by Clark et al. (2021). **(B)** In meta-communities, multiple species ("sp") form communities located at multiple locations ("loc") forming a region. We define the regional stability as the stability of the total biomass of one species across locations, and community stability as the stability of the total biomass across species in one location.

Previous studies have found that species diversity increases the invariability of communities (Thébault and Loreau, 2005; Tilman et al., 2006; Gross et al., 2014), usually as a consequence of the asynchrony of the population dynamics (Yachi and Loreau, 1999; Ives et al., 2000): variability decreases with species diversity because of the statistical averaging of the fluctuations in species' abundances for not perfectly synchronous dynamics (Doak et al., 1998). Hence, communities are more invariable than their constituent species, and the ratio of the community and species invariabilities has been proposed as an estimate of the across-species synchrony of the population dynamics (Loreau and De Mazancourt, 2008). This result can create differences between species- and community-scale stability assessments (Flöder and Hillebrand, 2012; Mougi and Kondoh, 2012; Downing et al., 2014).

Analogously, studies of meta-communities (defined as sets of local communities that are linked by dispersal of multiple potentially interacting species; Leibold et al., 2004) have also proven that temporal invariability increases from local species to meta-communities (Wang and Loreau, 2014), again as a consequence of decreasing synchrony (Wang et al., 2019). The ratio between regional and local invariabilities could be employed as a proxy for a region-wide synchrony, which would represent the global degree of synchrony in the spatial network (not to be confused with the regional synchrony that might occur at smaller spatial scales: Moran, 1953; Lande et al., 1999; Jarillo et al., 2018, 2020). Heterogeneous environmental conditions (Chesson, 2000) and the dispersal ability of the species (Amarasekare, 2008) might further cause the spatial scale to influence population and community dynamics, and therefore spatial scale-dependence of stability. Similarly, also the precise spatial organization of the network may influence meta-community stability, as has been found when comparing riverine vs linear networks (Fagan, 2002; Carrara et al., 2012; Altermatt, 2013; Liu et al., 2013; Peterson et al., 2013).

Finally, ecological and spatial scales may interact. For instance, spatial scale affects stability more in communities than in populations (Mougi and Kondoh, 2016), depending on the position of the focal species within the community (Limberger et al., 2019). Furthermore, decreasing the size of spatial networks reduces species richness more than one would expect from spatial samplings (Chase et al., 2020).

To understand the scaling of stability in meta-communities, which will allow the comparison of stability of systems analyzed at different scales, Clark et al. (2021) provided general scaling laws of common stability components: resistance, initial resilience, and invariance. They found that – if these measures are not normalized by biomass – invariance, resistance and initial resilience decrease with the spatial scale (the size of the considered spatial region). While they found that ecological scale (the number of species) also in most cases reduced invariance and resistance, it increased initial resilience. Because total biomass often changes with scale, we first wanted to revisit the scaling laws established by Clark et al. (2021) by considering normalized stability measures.

Our objective is to investigate whether normalized regional and community stability measures are simple summary statistics of local and species stability, respectively, and do not change with the spatial or ecological scale at which they are studied (Fig. 1B). Additive quantities (also known as extensive) are quantities that are simply added when combining several subsystems (IUPAC, 2019). They include the total biomass, its derivative, or its change due to a perturbation. The quotient of two additive quantities gives what is call an intensive quantity, which are scale-free quantities (independent of the number of subsystems) (IUPAC, 2019). Therefore, additive quantities automatically give scale-free quantities when normalized by another additive quantity like the number of subsystems or the total biomass.

We begin our paper (Section 2) by introducing biomass-normalized stability measures of growth rate, resistance, initial resilience, and invariability. In Section 3, we then show that resistance, growth rate, and initial resilience are all independent of scale. This is because the introduced normalized definitions of these stability components are intensive magnitudes, which do not scale with the size of the systems. In contrast, invariability is shown to be independent of scale only in the case of perfectly synchronous dynamics of species and local nodes (the different locations forming the spatial network). In the more realistic scenarios with at least some level of asynchrony, invariability is not an intensive quantity, and it increases with the network size. We then discuss how to use these results to statistically estimate regional/community stability from partial information (values on some local nodes or species). Section 4 shows that our formulas compare well with model simulations. Finally, Section 5 discusses various ecologically relevant aspects of the results: the influence of low-stability nodes or species on network stability, the relevance of the mathematical definitions, the implications for the empirical measurement of stability, and the implications for the stability-complexity debate.

## 2    Biomass normalized stability measures

We introduce biomass-normalized measures for the main stability components: resistance (to a perturbation), growth rate (after a perturbation), initial resilience, and invariance (Fig. 1A). These biomass-normalized measures avoid potential scale effects due to the usual scale-dependence of biomass. Biomass, its derivative and the change of biomass due to a perturbation are additive (extensive) quantities. Their quotients are expected to be scale-free measures (intensive magnitudes). We advance that for invariance the study will be more complicated as the variance is not additive.

### 2.1    Growth rate

We define the growth rate $R$ as the relative instantaneous return rate to the equilibrium of the biomass $N$ after any sudden biomass change caused by any external perturbation at time $t_0$,

$$R \equiv \frac{1}{N(t)} \frac{dN(t)}{dt} \bigg|_{t=t_0} . \tag{1}$$

Growth rate after a perturbation could be argued not to be a proper stability measure. Growth rate provides the rate of change of the population relative to the remaining population. Instead, initial resilience (section 2.3) provides the rate of change of the population relative to the departure of the population to its equilibrium value. This makes the initial resilience a more intuitive stability measure, as it estimates the initial rate of return to equilibrium. However, as we will show below, the growth rate is directly proportional to the scale-dependence of initial resilience, which shows that it contains information on stability and therefore can be considered a component of stability.

## 2.2 Resistance

We define the resistance $\Omega$ as the inverse of the relative change of biomass as a consequence of a perturbation (Isbell et al., 2015; Baert et al., 2016),

$$\Omega \equiv \frac{N(t_0)}{N(t_0) - N(t_0 + \delta t)} \; . \tag{2}$$

Instead of working with this measure, sometimes its inverse $\Omega^{-1}$ is referred to as resistance (Yang et al., 2019), with the possible conceptual disadvantage of presenting smaller values for more resistant systems. Other studies define resistance as the logarithm of the ratio of biomasses before and after any disturbance, $\ln\big(N(t_0 + \delta t)/N(t_0)\big)$ (Hillebrand et al., 2018), whose absolute value will also decrease as systems become more resistant. Actually, in absolute value this logarithmic definition is at first order equivalent to $\Omega^{-1}$ for small perturbations (as can be proven by applying a Taylor expansion on $|N(t_0 + \delta t) - N(t_0)|/N(t_0) \ll 1$), so its inverse is at first order equivalent to Eq. (2).

## 2.3 Initial resilience

We define the initial resilience as the initial rate at which a biomass perturbation disappears, normalized by the extent of the perturbation

$$\rho \equiv \frac{1}{|N(t) - N^*|} \frac{\mathrm{d}(N(t) - N^*)}{\mathrm{d}t}\bigg|_{t=t_0}, \tag{3}$$

where $N^*$ stands for the equilibrium biomass, which is assumed to be equal to the biomass just before the perturbation ($N(t_0)$). This definition stands for the short term recovery rate after a perturbation (Arnoldi et al., 2018), and has been sometimes referred to as reactivity (Neubert and Caswell, 1997). This initial resilience $\rho$ can be expressed in terms of the already considered growth rate $R$ (Eq. (1)) and resistance $\Omega$ (Eq. (2)),

$$\rho \equiv \frac{1}{|N(t) - N^*|} \frac{\mathrm{d}(N(t) - N^*)}{\mathrm{d}t}\bigg|_{t=t_0} = \frac{N^*}{|N(t) - N^*|} \frac{1}{N^*} \frac{\mathrm{d}N(t)}{\mathrm{d}t}\bigg|_{t=t_0} = \Omega\,R. \tag{4}$$

## 2.4 Invariability

We define invariability $I$ as the ratio of the square temporal mean of the biomass and its temporal variance (Thibaut and Connolly, 2013):

$$I \equiv \frac{\big[\mathrm{mean}_t\big(N(t)\big)\big]^2}{\mathrm{var}_t\big(N(t)\big)} \; . \tag{5}$$

This quantity is the inverse of the squared coefficient of variation of the biomass. Note that if the system is stable enough to stay away from extinction, this invariability will necessary be greater than 1; otherwise, environmental fluctuations might bring the biomass to zero. Other invariability estimates further normalize this invariability by the amplitude of environmental stochasticity (Haegeman et al., 2016; Arnoldi et al., 2019), in order to compare the invariability of systems subject to different environmental variability conditions.

## 3    Spatial and ecological scaling of stability measures

In this section, we address the spatial (local vs. regional) and ecological (species vs. community) scaling behavior (Fig. 1B) of the previously introduced stability measures (Section 2, Fig. 1A). Spatial scaling refers to how the measure changes from the local level (*e.g.*, one location) to the regional level (*e.g.*, all locations). Ecological scaling refers to how the measure changes from the species level to the community level (*e.g.*, all species). Knowing the response of stability metrics to scaling is important to build estimators that can be applied to the empirical study of extended ecological networks, which can only be partially sampled. We start with the study of the growth rate, as the simpler case, and follow with resistance, initial resilience, and invariability.

### 3.1    Growth rate

Using our definition of growth rate (Eq. (1)), we can compute the growth rate after a perturbation of a given species $i$ at a given specific location $x$, $R_{x,i}$, as the normalized time derivative of the species local biomass, $N_{x,i}$,

$$R_{x,i} = \frac{1}{N_{x,i}(t)} \frac{\mathrm{d}N_{x,i}(t)}{\mathrm{d}t} \bigg|_{t=t_0} . \tag{6}$$

Defining the regional biomass of one species as the sum of all local biomasses of that species across the spatial network, $N_i \equiv \sum_x N_{x,i}$, and based on the mathematical definition of growth rate (Eq. (1)) and on the sum rule of the derivative, we obtain that the regional growth rate $R$ of the species $i$ is

$$R_i^{(\mathcal{R})} \equiv \frac{1}{N_i(t)} \frac{\mathrm{d}N_i(t)}{\mathrm{d}t} \bigg|_{t=t_0} = \frac{\sum_x N_{x,i}(t_0) R_{x,i}}{\sum_x N_{x,i}(t_0)} , \tag{7}$$

meaning that the regional growth rate of the species is the weighted arithmetic mean of local species growth rates, $R_{x,i}$, with weights equal to the local species biomasses at the moment of the perturbation, $N_{x,i}(t_0)$ (Fig. 2). Analogously, the local community growth rate $R_x^{(\mathcal{C})}$, or the growth rate of the sum of biomasses across all the species of the community at a specific location $N_x \equiv \sum_i N_{x,i}$, is

$$R_x^{(\mathcal{C})} \equiv \frac{1}{N_x(t)} \frac{\mathrm{d}N_x(t)}{\mathrm{d}t} \bigg|_{t=t_0} = \frac{\sum_i N_{x,i}(t_0) R_{x,i}}{\sum_i N_{x,i}(t_0)} , \tag{8}$$

*i.e.*, the local community growth rate is the weighed arithmetic mean of local species growth rates of each of the species, $R_{x,i}$, with weights equals to the local species biomasses at the moment of the perturbation, $N_{x,i}(t_0)$ (Fig. 2). Finally, we can define regional community growth rate $R^{(\mathcal{R}\mathcal{C})}$ (equal to the regional growth rate of the community, or to the community growth rate of the spatial network), as the growth rate of the total biomass across species and locations, $N_T \equiv \sum_x \sum_i N_{x,i}$, which is given by

$$R^{(\mathcal{R}\mathcal{C})} \equiv \frac{1}{N_T(t)} \frac{\mathrm{d}N_T(t)}{\mathrm{d}t} \bigg|_{t=t_0} = \frac{\sum_x \sum_i N_{x,i}(t_0) R_{x,i}}{\sum_x \sum_i N_{x,i}(t_0)} = \frac{\sum_x N_x(t_0) R_x^{(\mathcal{C})}}{\sum_x N_x(t_0)} = \frac{\sum_i N_i(t_0) R_i^{(\mathcal{R})}}{\sum_i N_i(t_0)} . \tag{9}$$

Thus, the regional growth rate of a community after a perturbation, $R^{(\mathcal{RC})}$, is the arithmetic mean of local community growth rates, weighted by the local total biomass; or equivalently the community growth rate of a spatial network is the arithmetic mean of the regional species growth rates, weighted by the species regional biomass.

Generally speaking, the network growth rate $R_{net}$ of any ecological or spatial network is then given by the biomass-weighted arithmetic mean of the growth rates at the nodes (either representing the species for an ecological network, or the local nodes for a spatial network). $R_{net}$ can represent either the regional value in a spatial network, the community value in an ecological network, or even the regional community value; computed using Eq. (7), (8) or (9) respectively. This $R_{net}$ can be also expressed as

$$R_{net} = \mu_R \left(1 + \frac{\sigma_N}{\mu_N} \frac{\sigma_R}{\mu_R} c_{N,R}\right), \tag{10}$$

where $\mu_N \equiv \text{mean}(N)$ and $\mu_R \equiv \text{mean}(R)$ denote unweighted population arithmetic means of biomasses and growth rates, respectively, $\sigma_N \equiv \sqrt{\text{var}(N)}$ and $\sigma_R \equiv \sqrt{\text{var}(R)}$ are their standard deviations computed as the square root of the variances, and $c_{N,R}$ the normalized correlation between biomass and growth rate. (See Table 1 for all the mathematical definitions). Given that $-1 \leq c_{N,R} \leq 1$, the network growth rate $R_{net}$ can be greater or smaller than the unweighted mean of growth rates $\mu_R$ depending on the positive or negative correlations between the node biomasses and node growth rates.

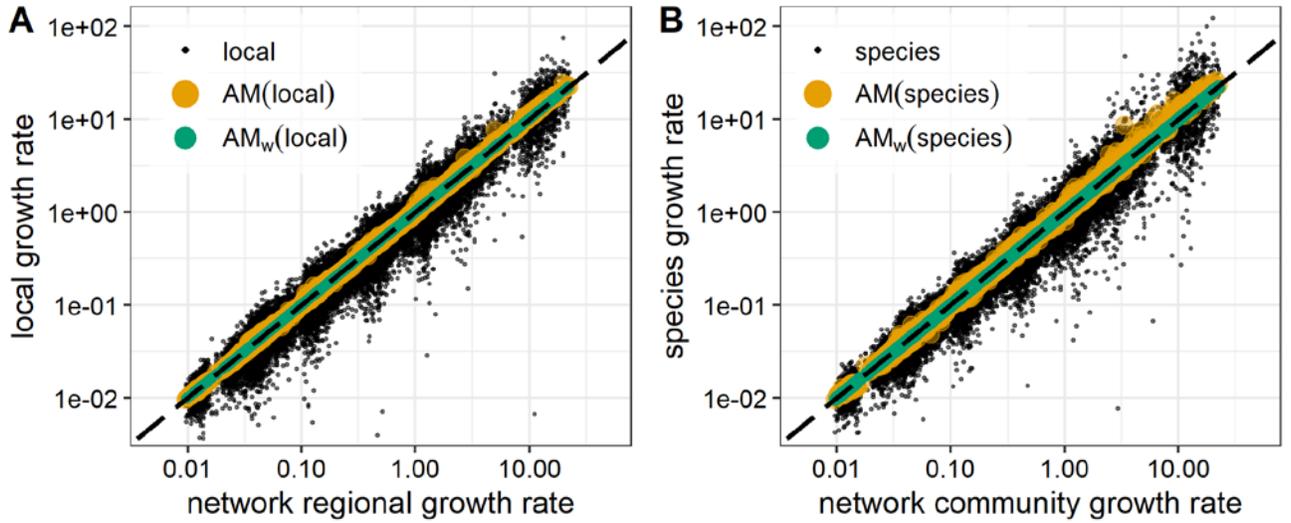

**Figure 2:** Regional (**A**) and community (**B**) growth rates, compared to local and species growth rates and to their unweighted arithmetic mean (AM) (yellow circles), and biomass weighted arithmetic mean (AM_w) (green circles). AM and AM_w values closer to the identity line (black, dashed) estimate more precisely the regional (panel A) and community growth rate (panel B). Black dots are the growth rates of individual localities (A) or species (B) on which the means AM and AM_w were computed. Data were generated for 6300 random communities of 10 competitors in 10-node random spatial networks (Fig. S1A and text on Supplementary Material), after a biomass decrease from the equilibrium affecting all species at all locations. AM_w are found to be better estimators of the growth rate, as expected from the results shown in the text.

| SYMBOL | DESCRIPTION |
|---|---|
| $N$ | Biomass. The equilibrium biomass is denoted as $N^*$. Moreover, the biomass of species $i$ at location $x$ and time $t$ is denoted as $N_{x,i}(t)$. |
| $R$ | Growth rate of the biomass after a perturbation, $R \equiv \frac{1}{N(t)}\frac{dN(t)}{dt}\Big|_{t=t_0}$. Measures of growth rate at regional ($R^{(\mathcal{R})}$), community ($R^{(\mathcal{C})}$), and regional community ($R^{(\mathcal{RC})}$) scales can be computed or estimated with weighted arithmetic means, Eqs. (7)-(9), or with Eq. (10). The error of the estimate is given by Eq. (11). |
| $\Omega$ | Resistance of the biomass to a perturbation, $\Omega \equiv N(t_0)/\big(N(t_0) - N(t_0 + \delta t)\big)$. Measures of resistance at regional ($\Omega^{(\mathcal{R})}$), community ($\Omega^{(\mathcal{C})}$), and regional community ($\Omega^{(\mathcal{RC})}$) scales can be computed or estimated with harmonic means (Eq. (13)), or with Eq. (14). The error or the estimate is given in Eq. (A5) in the Supplementary Material. |
| $\rho$ | Initial resilience of the biomass after a perturbation, $\rho \equiv \frac{1}{|N(t)-N^*|}\frac{d(N(t)-N^*)}{dt}\Big|_{t=t_0}$. Measures of initial resilience at regional ($\rho^{(\mathcal{R})}$), community ($\rho^{(\mathcal{C})}$), and regional community ($\rho^{(\mathcal{RC})}$) scales can be computed or estimated with the estimates of growth rate $R$ and resistance $\Omega$ (Eq. (16)). The error or the estimate can be also determined with the estimates of $R$ and $\Omega$ (see Appendix A3 of the Supplementary Material). |
| $I$ | Invariability of the biomass, defined as the inverse of the squared temporal coefficient of variation of the biomass, $I \equiv \frac{[\mathrm{mean}_t(N(t))]^2}{\mathrm{var}_t(N(t))}$. Measures of invariability at regional ($I^{(\mathcal{R})}$), community ($I^{(\mathcal{C})}$), and regional community ($I^{(\mathcal{RC})}$) scales can be computed or estimated with Eq. (22) (and Eq. (D14) of the Supplementary Material), and generally depends on the number of locations and species. The error of the estimate should be determined with bootstrapping techniques. |
| $\mu_X$ and $M_X$ | Population and sample unweighted means of variable $X$. The population mean is computed when all $n$ nodes of the network were measured, $\mu_X \equiv \frac{1}{n}\sum_{i=1}^{n}X_i$, while the sample mean is computed when just a sample of $\tilde{n} < n$ nodes are measured, $M_X \equiv \frac{1}{\tilde{n}}\sum_{i=1}^{\tilde{n}}X_i$. |
| $\sigma_X^2$ and $S_X^2$ | Population and sample variances of variable $X$: $\sigma_X^2 \equiv \frac{1}{n}\sum_{i=1}^{n}(X_i - \mu_X)^2$ and $S_X^2 \equiv \frac{1}{\tilde{n}}\sum_{i=1}^{\tilde{n}}(X_i - M_X)^2$. |
| $c_{X,Y}$ and $C_{X,Y}$ | Population and sample Pearson correlation coefficients of variables $X$ and $Y$: $c_{X,Y} \equiv \left[\frac{1}{n}\sum_{i=1}^{n}(X_i - \mu_X)\,(Y_i - \mu_Y)\right]/(\sigma_X\,\sigma_Y)$ and $C_{X,Y} \equiv \left[\frac{1}{\tilde{n}}\sum_{i=1}^{\tilde{n}}(X_i - M_X)\,(Y_i - M_Y)\right]/(S_X\,S_Y)$. Even though they are usually denoted as $\rho_{X,Y}$ and $r_{X,Y}$, the employed notation was preferred to avoid possible confusions with initial resilience. |

**Table 1:** Summary table: symbols used in the article and their descriptions.

Previous Eqs. (7)-(9) provide accurate computations of the regional, community or regional community growth rate if we know the growth rates of all involved nodes (either species or locations) (Fig. 2). However, in many practical situations, we only can sample a limited number of nodes, $n$. We can then estimate the network growth rate $\widetilde{R_{net}}$ using Eq. (10) with the sampled nodes, replacing the population means ($\mu_N$ and $\mu_R$) by the sample means ($M_N$ and $M_R$), the population standard deviations ($\sigma_N$ and $\sigma_R$) by the sample standard deviations ($S_N$ and $S_R$), and the population correlation ($c_{N,R}$) by the sample correlation ($C_{N,R}$) (see Table 1). *I.e.*, computing the biomass-weighted arithmetic mean of the sampled node growth rates. As the network estimate of growth rate corresponds to the weighed arithmetic mean of the node growth rates, we can estimate the standard error that arises from a partial (but representative) sampling using the formulas provided by Cochran (1977) and validated by Gatz and Smith (1995) (see Appendix A of the Supplementary Material)

$$\text{SE}\left(\widetilde{R_{net}}\right) = t_{\tilde{n}-1} \frac{S_R}{\sqrt{\tilde{n}-1}} \sqrt{1 + \left(1 - C_{N,R}^2\right)\left(\frac{S_N}{M_N}\right)^2 + C_{N,R}^2 \left(\frac{S_N}{M_N}\right)^4}, \qquad (11)$$

where $\text{SE}\left(\widetilde{R_{\text{net}}}\right)$ is the standard error of the network, corresponding to a 95% confidence level; $\tilde{n}$ is the sample size (*i.e.*, the number of sampled nodes); and $t_{\tilde{n}-1}$ is the Student's t distribution with $\tilde{n}-1$ degrees of freedom associated with a 95% confidence level, whose value is approximately 1.96 for large enough sampling sizes. Eq. (11) shows that the uncertainty in the determination of the regional community growth rate $\text{SE}\left(\widetilde{R_{\text{net}}}\right)$, as expected, decreases when sampling more localities or species (larger $\tilde{n}$), and when the growth rates vary less across localities or species (smaller $S_R$). Note that, generally, $S_N < M_N$, and so also biomasses that vary less relative to their average biomass will lead to less variable estimates of $R_{net}$. Another implication is that positive or negative correlations between biomass and growth rate will mostly decrease $\text{SE}\left(\widetilde{R_{\text{net}}}\right)$, because $S_N/M_N < 1$ and so $(S_N/M_N)^4 \ll (S_N/M_N)^2$. Nevertheless, since negative correlations decreases the network growth rate (Eq. (10)), negative correlations between $N$ and $R$ require larger sample sizes to control the relative error on the estimated network growth rate (Fig. 3).

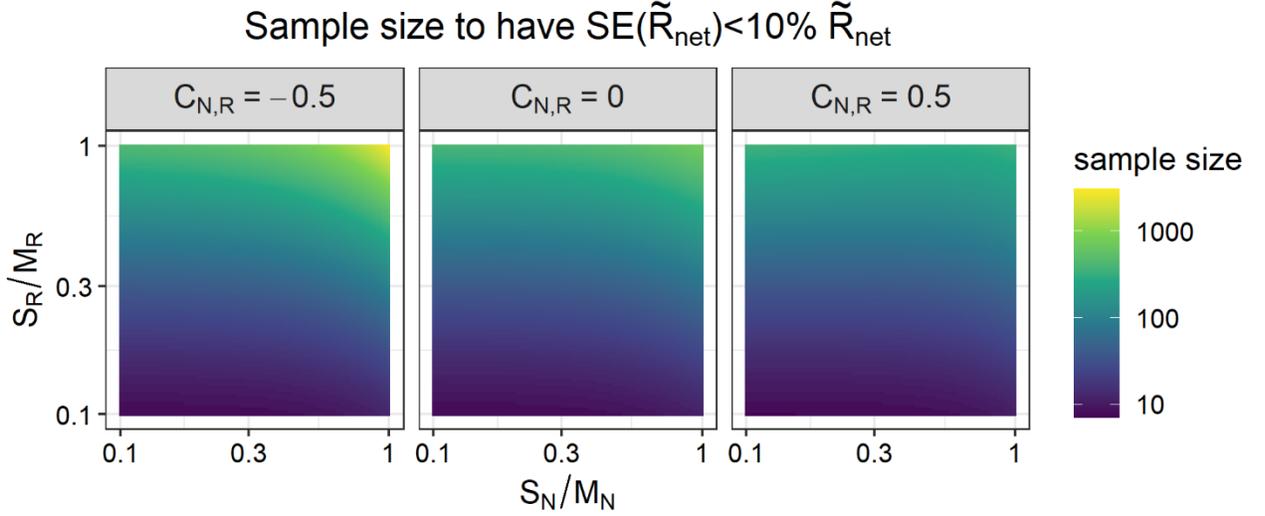

**Figure 3**: Sample size required to estimate the growth rate of an ecological or spatial network from estimates at the nodes such that the standard error of the estimate is smaller than 10%. The required sample size mainly increases with the coefficient of variation of the node estimates of growth rate ($S_R/M_R$), and to a lesser extent with the coefficient of variation of the node biomass ($S_N/M_N$). Larger sampling sizes are required for more negative values of the cross-correlation between the biomass and the growth rate $C_{N,R}$ (Table 1).

### 3.2   Resistance

The resistance of species $i$ at location $i$, defined as in Eq. (2), is

$$\Omega_{x,i} \equiv \frac{N_{x,i}(t_0)}{N_{x,i}(t_0) - N_{x,i}(t_0 + \delta t)} \ . \tag{12}$$

Then, from its definition, the regional and community resistances are the harmonic means of local or species resistances, weighted by the local or species biomasses (Appendix B, Fig. 4),

$$\Omega_i^{(\mathcal{R})} = \left[ \frac{\sum_x N_{x,i}(t_0) \frac{1}{\Omega_{x,i}}}{\sum_x N_{x,i}(t_0)} \right]^{-1} \ , \qquad \Omega_x^{(\mathcal{C})} = \left[ \frac{\sum_i N_{x,i}(t_0) \frac{1}{\Omega_{x,i}}}{\sum_i N_{x,i}(t_0)} \right]^{-1} \ . \tag{13}$$

Hence, for any ecological or spatial network the network resistance is the weighted harmonic mean of node resistances, weighted by the node biomasses (see Appendix B). Again, network resistance can be also rewritten in a more general way as

$$\Omega_{net} = \frac{1}{\mu_{\Omega^{-1}}} \frac{1}{1 + \frac{\sigma_N}{\mu_N} \frac{\sigma_{\Omega^{-1}}}{\mu_{\Omega^{-1}}} c_{N,\Omega^{-1}}}, \tag{14}$$

where again $\mu$ represents unweighted arithmetic means, $\sigma$ the population standard deviations, and $c_{N,\Omega^{-1}}$ the correlation between biomasses and the inverse of resistance (see Table 1). Hence, it is possible also to estimate the network resistance from a partial sampling of the network, replacing in Eq. (14) the population means, standard deviations and correlations by their sample equivalents.

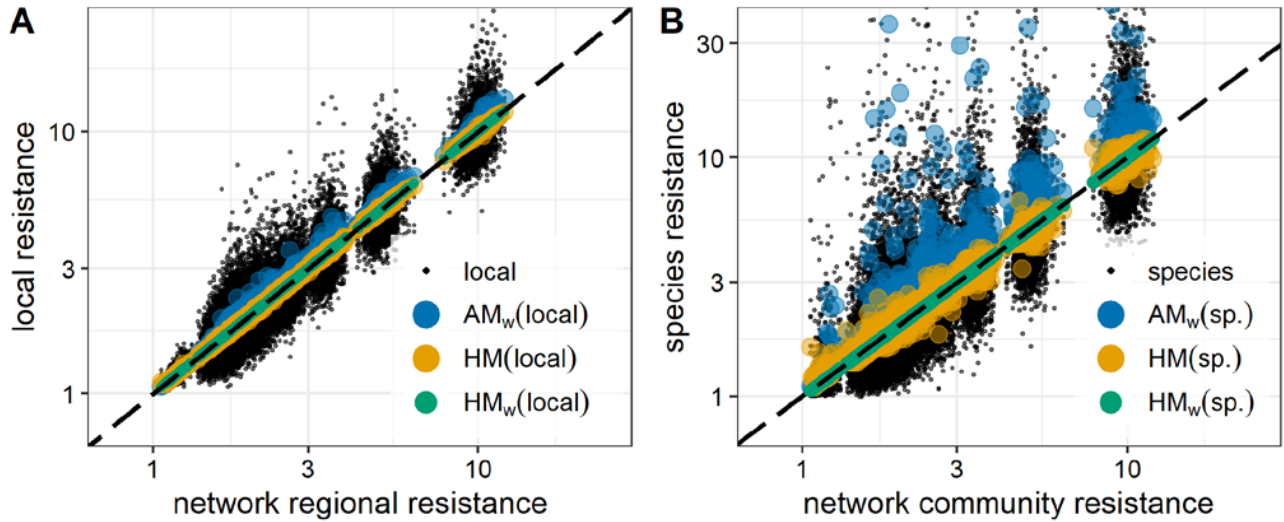

**Figure 4:** Regional (**A**) and community (**B**) resistances, compared to local and species resistances and to their biomass weighted arithmetic mean ($AM_w$) (blue circles), unweighted harmonic mean (HM) (yellow circles), and biomass weighted harmonic mean ($HM_w$) (green circles). $AM_w$, HM, and $HM_w$ values closer to the identity line (black, dashed) estimate more precisely the regional (panel A) and community resistance (panel B). Black dots are the resistances of the individual localities (A) or species (B) on which the means were computed. Data were generated for 6300 random communities of 10 competitors in 10-node random spatial networks (Fig. S1A and text on Supplementary Material), from the biomass decrease caused by a reduction of the species growth rates at all locations. The harmonic mean of resistances is always smaller than the arithmetic mean (see Appendix C of the Supplementary Material). Moreover, $HM_w$ are found to be better estimators for the resistance, as expected from the results shown in the text.

Even though the resistance of a network $\Omega_{net}$ is the weighted harmonic mean of resistances, the network estimate of its inverse $(\Omega^{-1})_{net}$ is the weighted arithmetic mean of the sub-units estimates of $\Omega^{-1}$. This result again allows us to estimate its standard error arising from incomplete but representative network sampling (Appendix A of the Supplementary Material), which let us to obtain an expression for the standard error of the resistance obtained from a partial sampling of a network (appendix B, Eq. B8). The relative uncertainty of the network resistance will be dominated by the number of samples from the network, and by the variance of the inverse of resistances.

### 3.3 Initial resilience

In Eq. (4), we show that initial resilience $\rho$ is given by the product of resistance $\Omega$ and growth rate $R$, *i.e.*, $\rho = \Omega\,R$. Therefore, to estimate $\rho$ for different scales one can use the already obtained scaling of $\Omega$ and $R$ (which are scale invariant and have the simple estimators described). For example, defining the local species initial resilience as

$$\rho_{x,i} \equiv \frac{1}{\left|N_{x,i}(t) - N_{x,i}^*\right|} \frac{d\left(N_{x,i}(t) - N_{x,i}^*\right)}{dt}\bigg|_{t=t_0}, \tag{15}$$

and by defining the regional species initial resilience as the initial resilience of the regional biomass of a given species, it is easy to see that it coincides with the product of the regional resistance and the regional growth rates

$$\rho_i^{(\mathcal{R})} \equiv \frac{1}{\left|N_i(t) - N_i^*\right|} \frac{d(N_i(t) - N_i^*)}{dt}\bigg|_{t=t_0} = \Omega_i^{(\mathcal{R})} R_i^{(\mathcal{R})}. \tag{16}$$

That is, we can express the regional initial resilience as the product of the regional estimates of resistance (the weighted harmonic mean of local resistances) and growth rate (the weighted arithmetic mean of local growth rates). An analogous result links the community initial resilience to the community estimates of resistance and growth rate. In particular, since both resistance and growth rates were scale invariant (the network estimates of $R$ and $\Omega$ were respectively the harmonic and arithmetic mean of the node estimates), also the initial resilience would be scale invariant. Actually, the regional initial resilience can be also expressed as

$$\rho_i^{(\mathcal{R})} = \frac{\sum_x N_{x,i}^* \Omega_{x,i}^{-1} \rho_{x,i}}{\sum_x N_{x,i}^* \Omega_{x,i}^{-1}}. \tag{17}$$

Thus, the network estimate of initial resilience is the arithmetic mean of the node estimates of initial resilience, weighted by the change of the node biomass caused by a perturbation, $N \Omega^{-1}$ (Fig. 5). This result reinforces that initial resilience is another scale invariant stability property of ecosystems, and that less resistant nodes with higher biomass will disproportionally influence the initial resilience of the total network. Again, for any spatial or ecological network, we can express the network estimate of the initial resilience in a general way, as the product of the network resistance (the weighted harmonic mean of node resistances, highly influenced by nodes with higher biomasses and lower resistances) and the network growth rate:

$$\rho_{net} = \Omega_{net} R_{net} = \mu_\rho \left( 1 + \frac{\frac{\sigma_N}{\mu_N} \frac{\sigma_\rho}{\mu_\rho} c_{N,\rho} + \frac{\sigma_\rho}{\mu_\rho} \frac{\sigma_{\Omega^{-1}}}{\mu_{\Omega^{-1}}} c_{\rho,\Omega^{-1}}}{1 + \frac{\sigma_N}{\mu_N} \frac{\sigma_{\Omega^{-1}}}{\mu_{\Omega^{-1}}} c_{N,\Omega^{-1}}} \right). \tag{18}$$

This result indicates that correlations between the node biomass, resistances and initial resiliences can make the network initial resilience higher or smaller than the unweighted mean of node initial resilience estimates.

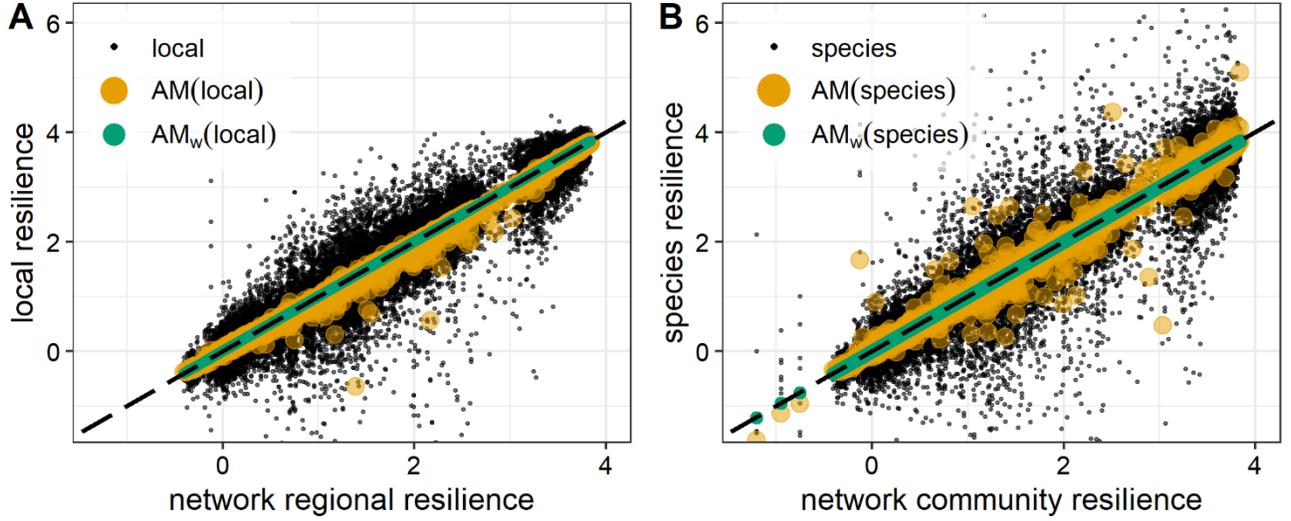

**Figure 5:** Regional (**A**) and community (**B**) initial resiliences, compared to local and species initial resiliences and to their unweighted arithmetic mean (AM) (yellow circles) and biomass weighted arithmetic mean (AM$_w$) (green circles) (with weights given by the difference between the actual and the equilibrium biomasses, Eq. (17)). AM and AM$_w$ values closer to the identity line (black, dashed) estimate more precisely the regional (panel A) and community initial resilience (panel B). Black dots are the initial resiliences of the individual localities (A) or species (B) on which the means were computed. Data were generated for 6300 random communities of 10 competitors in 10-node random spatial networks (Fig. S1A and text on Supplementary Material), after a biomass decrease from the equilibrium affecting all species at all locations. AM$_w$ are found to be better estimators of the initial resilience, as expected from the results shown in the text.

As for growth rate and resistance, we can also estimate network initial resilience from an incomplete sampling of the network as $\widetilde{\rho_{net}} = \widetilde{\Omega_{net}} \, \widetilde{R_{net}}$. And assuming no correlations between resistance and growth rates, the standard error committed with such sampling would be $\text{SE}(\widetilde{\rho_{net}}) = \widetilde{\Omega_{net}} \, \text{SE}(\widetilde{R_{net}}) + \widetilde{R_{net}} \, \text{SE}(\widetilde{\Omega_{net}})$. In such a case, the relative uncertainty of the estimated initial resilience would be first given by the sample size, and then by the variances of the growth rates and reciprocal resistances.

### 3.4 Invariability

We consider the invariability definition of Eq. (5), so the local species invariability reads

$$I_{x,i} \equiv \frac{\left[\text{mean}_t\left(N_{x,i}(t)\right)\right]^2}{\text{var}_t\left(N_{x,i}(t)\right)} \,. \tag{19}$$

The regional invariability can be defined as the invariability of the total biomass of one species across all locations in a spatial network. For the synchronous space ("ss") case, for which the local biomass dynamics are perfectly positively correlated, the regional invariability of species $i$ reads (see appendix D)

$$I_i^{(\mathcal{R};ss)} = \left[\frac{\sum_x N_{x,i}^*}{\sum_x N_{x,i}^* \frac{1}{\sqrt{I_{x,i}}}}\right]^2 .$$

(20)

Then, for perfectly synchronous local dynamics, the regional species invariability is the square of the harmonic mean of the square root of the local species invariabilities, weighted by the equilibrium local biomass densities. Conversely, if the local species biomass dynamics is spatially asynchronous (*asynchronous space,* "as"), the regional species invariability of the whole spatial network is

$$I_i^{(\mathcal{R};as)} = \frac{\left(\sum_x N_{x,i}^*\right)^2}{\sum_x \left(N_{x,i}^*\right)^2 \frac{1}{I_{x,i}}} = n_L \; \frac{1}{1 + \left(\frac{\sigma_{N_i^*}}{\mu_{N_i^*}}\right)^2} \; \frac{\sum_x \left(N_{x,i}^*\right)^2}{\sum_x \left(N_{x,i}^*\right)^2 \frac{1}{I_{x,i}}} .$$

(21)

For the asynchronous-space case, the regional invariability is proportional to the number of locations $n_L$ (Fig. 6). I.e., for the asynchronous-space case, invariability would be an extensive stability property, that grows linearly with the size of the system. In this asynchronous case, invariability is also proportional to the harmonic mean of local species invariabilities weighted by the squared local species biomasses. In addition, it is modulated by the spatial variance $\sigma_{N_i^*}^2$ and mean $\mu_{N_i^*}$ of the local equilibrium biomasses of species $i$. It can be proven that invariability is higher for asynchronous than for synchronous dynamics (see appendix D). Moreover, the number of locations does not modify the local invariability estimates, and there are not significative differences between cases with synchronous or asynchronous dynamics (Fig. 6B)

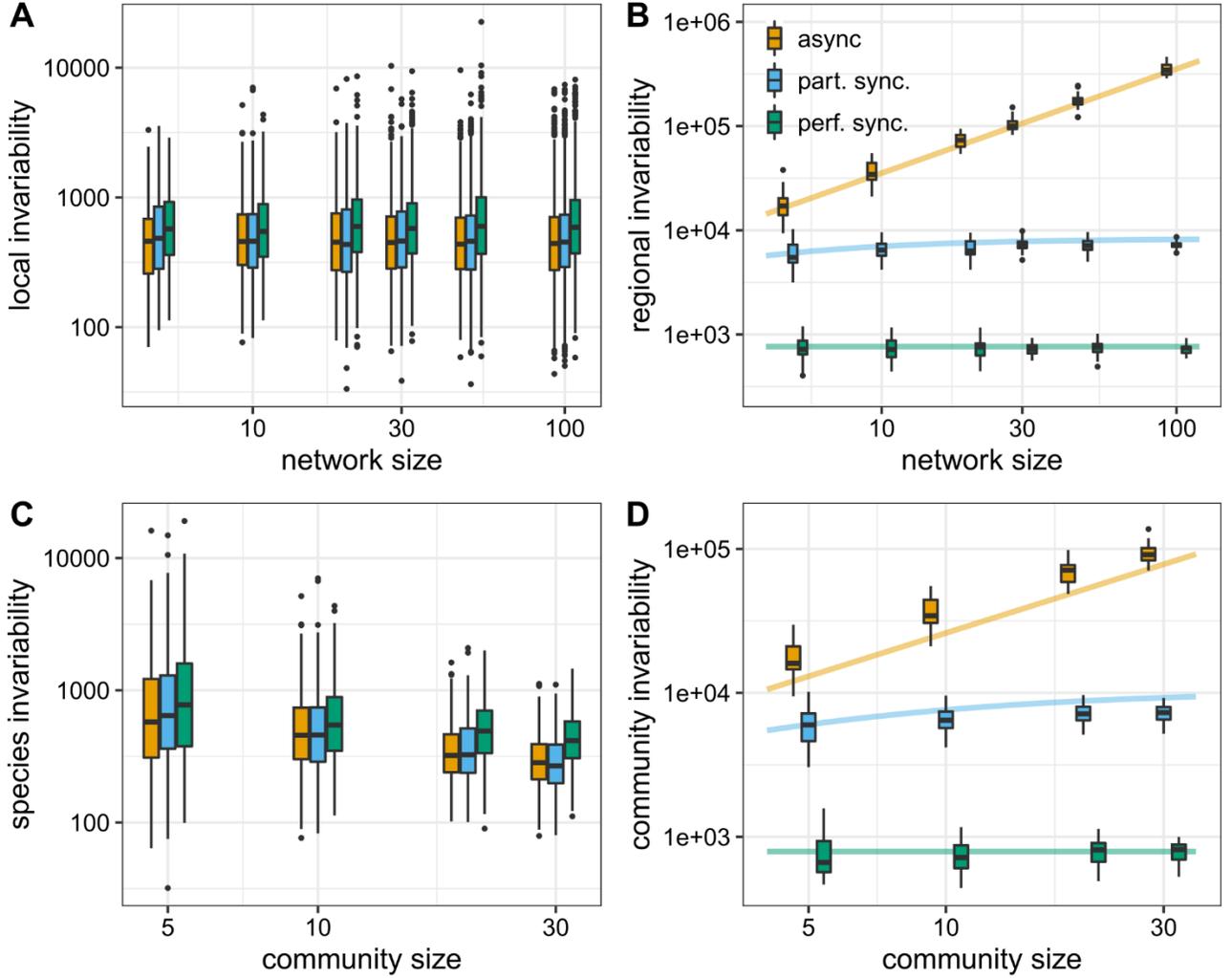

**Figure 6:** Local (**A**) and regional (**B**) invariability estimates in random spatial networks of random communities of 10 competitor species, for different sizes of the spatial network; and species (**C**) and community (**D**) invariability estimates of random communities of competitors at 10-node random spatial networks, for different number of species forming the communities. In panels A and B, we have considered three different scenarios: asynchronous local dynamics ($\bar{c} = 0$, yellow box plots), partially-synchronous local dynamics ($\bar{c} = 0.36$, blue), and perfectly-synchronous local dynamics ($\bar{c} = 1$, green). In panels C and D, we have considered the cases of asynchronous ($\tilde{c} = 0$), partially-synchronous ($\tilde{c} = 0.17$) and perfectly-synchronous ($\tilde{c} = 1$) species dynamics. Solid lines depict the invariabilities predicted by analytical expressions (Eqs. (20)-(22), and analogous expressions for community invariability).

The more general case of not perfectly synchronous dynamics can be expressed as

$$I_i^{(\mathcal{R})} = \left[ (1 - \bar{c}_i) \frac{1}{I_i^{(\mathcal{R};as)}} + \bar{c}_i \frac{1}{I_i^{(\mathcal{R};ss)}} \right]^{-1} . \tag{22}$$

where $\bar{c}_i$ is the typical correlation between different locations (Eq. (D10) of the appendix D). For the case $\bar{c}_i > 0$, and since by definition $\bar{c}_i \leq 1$, $I_i^{(\mathcal{R})}$ would simply be the weighted harmonic mean of $I_i^{(\mathcal{R};as)}$ and $I_i^{(\mathcal{R};ss)}$, with weights equal to $1 - \bar{c}_i$ and $\bar{c}_i$, respectively. And since, even though $I_i^{(\mathcal{R};ss)}$ does not depend on the number of locations $n_L$ (Eq. (20)), $I_i^{(\mathcal{R};as)}$ increases with $n_L$ (Eq. (21)), the resulting regional species invariability would increase as well with $n_L$ (except for the special case $\bar{c}_i = 1$). For the case $\bar{c}_i < 0$, since $I_i^{(\mathcal{R};as)} > I_i^{(\mathcal{R};ss)}$, the regional species invariability $I_i^{(\mathcal{R})}$ would be larger than $I_i^{(\mathcal{R};as)}$. Since $I_i^{(\mathcal{R};as)}$ increases linearly with the number of locations, the regional species invariability would then also increase with the number of locations. In summary, when the local population dynamics are not perfectly synchronized (so the typical spatial correlation of the local biomasses $\bar{c}$ is less than 1), the regional invariability increases with the number of locations of the spatial network (Fig. 6).

For community invariability, we can obtain completely analogous expressions to Eqs. (20)-(22), In particular, this proves that community invariability increases with the number of species forming the community, except for the special case of perfectly synchronous dynamics across species (Fig. 6C), while the degree of synchrony and the number of species do not significantly affect the invariabilities at the species level (Fig. 6D).

In general, network invariability is not a mean of the invariability estimates at the network nodes, so we cannot estimate its standard error in the same way that we did for resistance, growth rate and initial resilience (Appendix A of the Supplementary Material). We did not pursue here the characterization of such network invariability standard error. To estimate the error that arises from incomplete network sampling, general bootstrapping techniques should be applied instead (Efron and Tibshirani, 1985; Hesterberg, 2011).

## 4    Model simulations

In this study, we have investigated how different stability components such as growth rate, initial resilience, resistance, and invariability scale from the local or species level to the regional or community level. We now compare these scaling laws to numerically simulated population dynamics of a community of 10 competitors with the Lotka-Volterra model (see Appendix E of the Supplementary Material) in 10-node random spatial networks (Fig. S1A of the Supplementary Material) (Figs. 2, 4-6). To ensure that the results do not depend on the chosen network, and motivated by fundamental differences of meta-community stability between linear and riverine networks (Fagan, 2002; Carrara et al., 2012; Altermatt, 2013; Liu et al., 2013; Peterson et al., 2013), we complement these results with results for realistic riverine dendritic networks (Fig. S1B of the Supplementary Material) generated in R (R Core Team, 2020) with the OCNet package (Carraro et al., 2020) (Figs. S2-S5 of the Supplementary Material). All simulations of community dynamics were done in Python 3.7 (Python Core Team, 2019).

The simulation results confirm our theoretical prediction that growth rate and initial resilience are scale-free stability properties, where regional and community estimates equal to the weighted arithmetic mean of the estimates at the local or species level (Figs. 2 and 5, Figs. S2 and S4 of the Supplementary Material). Also, the simulations confirm that resistance is another scale-free property: the regional and community estimates of resistance are the harmonic mean of the local and species resistance estimates, weighted by the local biomasses or the species proportions (Fig. 4, Fig. S3 of the Supplementary Material). The numerical simulations also confirmed that invariability is a scale-free property solely in networks with perfectly synchronous dynamics for which all sub-units effectively act as a unique single unit (species or location). In more realistic networks, with imperfect synchrony across subunits, the invariability is higher than for the perfectly synchronous case (Fig. 6, Fig. S5 of the Supplementary Material), and it increases with the network size, so the regional or community invariability is actually larger than the average of its elements, and this difference is more pronounced in larger networks. Thus, realistic spatial networks are more invariable than their individual locations, and community dynamics are more invariable than the population dynamics of the species forming the community (Loreau and De Mazancourt, 2008; Haegeman et al., 2016).

## 5    Discussion

We have shown that resistance and initial resilience (and growth rate) of ecological or spatial networks, unlike invariability, are biomass-weighted means of the estimates of these stability measures at the nodes of the network. In this section, we will discuss the consequences of this fundamental difference between these stability components.

### 5.1    Resistance and initial resilience are scale-free network properties, while invariability is not

Some stability components, such as invariability, have been found to increase with the ecological (Thébault and Loreau, 2005; Tilman et al., 2006; Gross et al., 2014) and spatial scale (Wang and Loreau, 2014; Wang et al., 2019), as a consequence of not perfectly synchronous dynamics among species and among locations (Doak et al., 1998). Hence, communities and regions are more invariable than their constituent species and locations. On the contrary, other stability components seem to not depend on scale (Haegeman et al., 2016). To solve this issue, Clark et al. (2021) have proposed different scaling laws of three different common stability components: resistance, initial resilience, and invariance.

Our analysis confirms that regional and community invariability is larger than local and species invariability, and generally increases with the size of the studied network (Fig. 6). Similar results were obtained by Wang and Loreau (2014), who showed that the regional variability decreases with the species richness and the region size. However, and as is the case for asymptotic resilience (Haegeman et al., 2016), network resistance and short-term or initial growth rate and resilience (independent of asymptotic resilience, and a better proxy of resilience in experiments (Arnoldi et al., 2016, 2018)) is the mean of the local and species values (Figs. 2, 4-5). As this result is a consequence of the mathematical definition of these stability components, it will hold for any spatial and ecological network of any complexity, and for any meta-community dynamics model (Figs. S2–S5 of the Supplementary Material).

These results contribute to a better understanding of the multidimensional nature of ecological stability. While stability properties can be correlated (Donohue et al., 2013), depending on the characteristic of the environmental fluctuations affecting the systems (Arnoldi et al., 2019; Radchuk et al., 2019), their scaling laws can introduce another axis of fundamental differentiation between different stability components. Indeed, one could distinguish between network-level stability components (those fundamentally depending on the topology and size of the ecological network) and node-level stability components (those reflecting network averages of the node-level estimates). As a consequence, the analysis of multiple components of stability of the ecosystems might be preferred to the employment of single metrics that aim to reproduce the whole ecosystem stability (Lemoine, 2020).

## 5.2 Resistance is more affected than initial resilience by the presence of low-stable nodes or species

Although both resistance and initial resilience are scale-free properties, they differ in how the network estimate is averaged from the node measures, which has important ecological consequences. Harmonic means are more affected by the presence of low numbers, and less affected by the presence of high numbers, than arithmetic means (Ferger, 1931). Hence, the presence of less resistant species and locations will affect network-level resistance much more than network-level initial resilience (see Appendix C of the Supplementary Material). High resilient nodes can easily compensate low resilient nodes, as such bolstering network initial resilience. This will lower the impact of stressors that only affect a fraction of the nodes (Supp and Ernest, 2014). However, this is not the case for resistance (Fig. 7). Low-resistant nodes limit the resistance of a network much more, which makes resistance a stability component that is more difficult to protect in ecological networks. For example, in spatially heterogeneous meta-populations, meta-community resistance will be mainly determined by the resistance of the less stable regions, while the meta-community initial resilience will be mostly given by the average spatial conditions. Thus, the heterogenous presence of stressors in the meta-community (McCluney et al., 2014) is expected to have a stronger effect on the network resistance than in the network initial resilience.

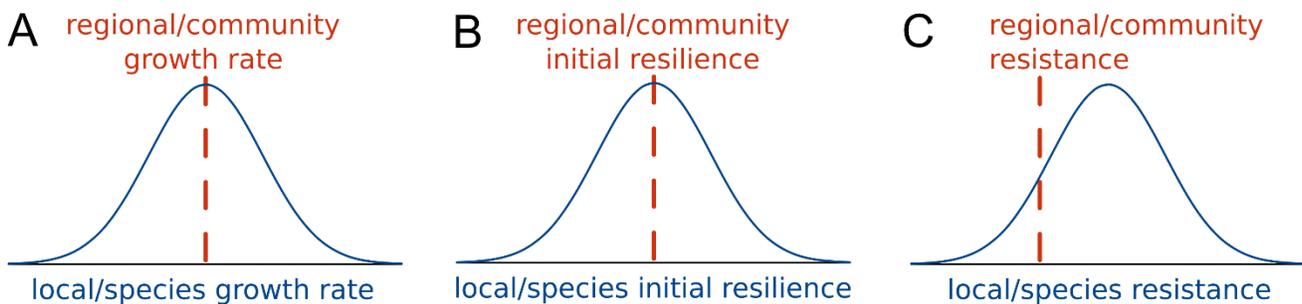

**Figure 7**. Schematic comparison of different stability components at local or species level vs at regional or community level, assuming normal distributions for the local and species estimates. The regional or community growth rate (**A**) and initial resilience (**B**) is the weighted arithmetic mean of the estimates at the local or species levels. On the contrary, regional/community resistance (**C**) is the weighted harmonic mean of the local/species estimates, so locations or species with low stability will limit the resistance of spatial or ecological networks.

## 5.3 Influence of mathematical definitions of stability

In this study, we have shown how different stability components scale from the local and species level to the regional and community level. Starting from common mathematical definitions, we showed that resistance and initial resilience are scale-free properties, while regions and communities are fundamentally more invariable than local species population dynamics. However, we anticipate that this result will depend on the employed mathematical definition (and then, on the proposed measurements) for these stability properties.

As previously noted, there is evidence that communities and spatial networks are more invariable than local species populations, as a consequence of imperfect synchronization on the local population dynamics (Gross et al., 2014; Wang et al., 2019). In this article, we have obtained the same result:

except for the unrealistic case of perfect inter-species and inter-location synchrony, meta-communities are more invariable than local populations, and such meta-community invariability increases with the number of species and the size of the spatial network (Fig. 6). Clark et al. (2021) have shown that invariance (the inverse of the biomass variance, not normalized by the average species biomasses) decreases with scale. Here, we show that invariability (i.e. biomass-normalized) increases with scale. Repeating our analyses using invariance recovers Clark et al.'s results (Appendix F of the Supplementary Material). Overall, this shows that normalizing the variance by the mean biomass (which increases with the network size) has a large influence on how we appreciate the scaling of stability. As networks will often be less variable than their nodes, we advocate the use of normalized stability properties related to variability when studying the effect of scale on this kind of stability. A similar difference between our results and those by Clark et al. (2021) on the scale dependence of resistance can be also explained from biomass normalization (Appendix G of the Supplementary Material).

For initial resilience, Clark et al. (2021) already employed a normalized definition. Here, with an analogous definition, we showed that the network initial resilience is the arithmetic mean of local species initial resiliences, being independent of the network size (Fig. 5). Clark et al. also found that the median of the initial resiliences was proportional to the ratio of the expected values of the invariance and the resistance for linear models. Since their defined invariance and resistance decrease with the characteristic ecological or spatial scale, their scale dependencies cancel out. An interesting question beyond the purpose of the present work is to what extent this relation between stability measures holds in the non-linear case.

The different means and behavior between resistance and initial resilience, discussed in section 5.2, depend on their mathematical definition and on the distribution of those stability components. For example, instead of resistance $\Omega$ (inverse of the relative change in biomass after a perturbation) we can define an alternative stability measure just given by the relative change of the biomass after a perturbation, *i.e.*, $\Omega^{-1}$. More resistant systems present smaller values of $\Omega^{-1}$, and $\Omega^{-1}$ represents the plasticity of the system against perturbations. Since the harmonic mean of a random variable is the inverse of the arithmetic mean of the reciprocals, it is easy to prove that a network estimator of $\Omega^{-1}$ would simply be the weighted arithmetic mean of the estimates at the nodes. For this new defined resistance, the presence of outliers affects the network resistance in the same way than the presence of outliers affected the network initial resilience, so nodes with above-average values of $\Omega^{-1}$ can be easily compensated by nodes with below-average values of $\Omega^{-1}$, having a limited effect on the network-level estimate of $\Omega^{-1}$. This is a clear indication of how the heterogeneous distribution of local species estimates (particularly its skewness (Stevens, 1955)) affects the network resistance and initial resilience.

The scale-free property found for the growth rate $R$, the resistance $\Omega$, and the initial resilience $\rho$ is due to their character as intensive quantities. The total biomass $N$, its derivative $\frac{dN}{dt}$, and the change in biomass due to a perturbation $N(t_0) - N(t_0 + \delta t)$, are are additive for subsystems. Their quotients have allowed us to construct quantities independent of the extent of the system, *i.e.*, scale-free quantities. Namely, growth rate $R$, resistance $\Omega$, and initial resilience $\rho$. For the simpler cases, the growth rate $R$ and the inverse of the resistance $\Omega^{-1}$ are given by a biomass-weighted arithmetic mean, which compensates the total biomass increase as the considered scale increases. The expression of $\Omega$ as a biomass-weighted harmonic mean is equivalent to the expression of $\Omega^{-1}$ as a biomass-weighted arithmetic mean and conserves the scale-free properties. The scale-free property of initial resilience $\rho$ can then be seen as a consequence of being the product (or quotient) of two intensive (or scale-free) quantities.

This view also shows why, in general, invariability is not scale-free. The temporal variance $\text{var}_t(N(t))$ is not extensive, because $\text{var}_t(N) = E_t[(N - E_t[N])^2] = E_t[N^2] - (E_t[N])^2$ is not additive for subsystems. Neither $\sqrt{\text{var}_t(N)}$ is extensive in general. This makes that only for completely synchronous dynamics the invariability is scale-free, as previously shown.

### 5.4 Implications for measuring stability empirically

Resistance and initial resilience of ecological spatial networks are biomass-weighted means of the local species estimates at the nodes of the networks, so they can be easily estimated from partial samples of the network. This property is important for the assessment of stability in large experiments (De Raedt et al., 2019; Karakoç et al., 2020; Saade et al., 2020) or field campaigns. Using our equations for the relative standard error of these stability indices (Eq. (11), and Eq. (A5) of Appendix A), one is able to estimate the sampling size required to control the error committed in the estimation of the network stability components from partial network samplings (illustrations in Fig. 3, and Fig. S6 of the Supplementary Material). The coefficient of variation of the studied stability property (resistance or initial resilience) affects this required sampling effort most (Fig. 3). Thus, the coefficient of variation will be higher for networks with more variable stability. Two other factors influence this variation: (1) a greater coefficient of variation of the biomasses; (2) a negative correlation between the biomass and the stability. (See Fig. S6 of the Supplementary Material.)

With respect to invariability, the standard error associated with an incomplete sampling is more difficult to estimate, since generally the network invariability is not a mean of the nodes' invariabilities, and depends on the size of the network. Hence, for this stability component the standard error should generally be assessed directly with a bootstrap. Moreover, for controlling the error associated with the estimation of network invariability from node-level invariability, it would be important to have an unbiased estimate of network size.

### 5.5 Implications for the stability-complexity debate

The stability-complexity debate (McCann, 2000; Allesina and Tang, 2015) originated from the disagreement between experimental observations often finding more complex systems to be more stable (Ives and Carpenter, 2007), and theoretical analyses finding more complex systems to be less stable (May, 1972; Pimm, 1984). To solve this disagreement, some authors have proposed generalizations of the original work of May (1972) that account for non-random among-species interactions (Yodzis, 1981; Rooney et al., 2006), or the stabilizing role of dispersal and spatial heterogeneity (Plitzko and Drossel, 2015; Gravel et al., 2016). Other approaches have suggested that the disagreement is caused by a focus on asymptotic resilience in theoretical studies (Pimm, 1984; McCann, 2000; Saeedian et al., 2021), which is biased by rare species (Haegeman et al., 2016). Our results adhere to this point of view, by showing that more complex systems (*i.e.* ecological and spatial networks, as opposed to single species and locations) are inevitably less variable, if using normalized estimates that correct for inherent effects on system size of system complexity. Consequently, such correction for system size leads to no relationship between complexity and the other stability properties (resistance and initial resilience). Taken together, these findings confirm that using a sole stability component (*e.g.* as asymptotic resilience) does not fully capture the complex ways in which biological systems deal with environmental changes (Pennekamp et al., 2018; Arnoldi et al., 2019). Assessing stability from a multi-dimensional perspective (Donohue et al., 2013; Arnoldi et al., 2019; Radchuk et al., 2019) will provide a more comprehensive picture and can reconcile apparent contradictions between and among theoretical and empirical studies.

## 6    Conflict of Interest

The authors declare that the research was conducted in the absence of any commercial or financial relationships that could be construed as a potential conflict of interest.

## 7    Author Contributions

JJ and FDL conceived the presented idea. JJ performed the numerical simulations and the analytical computations. FJCG and FDL verified the analytical derivations. JJ wrote the first draft and prepared the figures. All authors discussed the results and contributed to the final manuscript.

## 8    Funding

This work was supported by CEFIC under LRI project ECO50, and by the special research fund (FSR) from UNamur. Computational resources have been provided by the Consortium des Équipements de Calcul Intensif (CÉCI), funded by the Fonds de la Recherche Scientifique de Belgique (F.R.S.-FNRS) under Grant No. 2.5020.11 and by the Walloon Region. FJCG was funded by European Regional Development Fund (ERDF) and by the Spanish Ministry of Economy and Competitiveness through Grant RTI2018-095802-B-I00, and by European Union's Horizon 2020 through grant agreement No 817578 TRIATLAS.

## 9    Acknowledgments

Authors acknowledge constructive discussions on these results with Jeff Arnoldi, Bart Haegeman, Michel Loreau and other scientists of the Ecological Station of Moulis (CNRS, France). FJCG acknowledges the warm welcome at the Ecological Station of Moulis (CNRS, France). All the authors acknowledge the reviewers for their valuable feedback and efforts towards improving this manuscript.

## 10    Bibliography

Allesina, S., and Tang, S. (2015). The stability–complexity relationship at age 40: a random matrix perspective. *Popul. Ecol.* 57, 63–75. doi:10.1007/s10144-014-0471-0.

Altermatt, F. (2013). Diversity in riverine metacommunities: A network perspective. *Aquat. Ecol.* 47, 365–377. doi:10.1007/s10452-013-9450-3.

Amarasekare, P. (2008). Spatial Dynamics of Foodwebs. *Annu. Rev. Ecol. Evol. Syst.* 39, 479–500. doi:10.1146/annurev.ecolsys.39.110707.173434.

Arnoldi, J.-F., Bideault, A., Loreau, M., and Haegeman, B. (2018). How ecosystems recover from pulse perturbations: A theory of short- to long-term responses. *J. Theor. Biol.* 436, 79–92. doi:10.1016/j.jtbi.2017.10.003.

Arnoldi, J. F., Loreau, M., and Haegeman, B. (2016). Resilience, reactivity and variability: A mathematical comparison of ecological stability measures. *J. Theor. Biol.* 389, 47–59. doi:10.1016/j.jtbi.2015.10.012.

Arnoldi, J. F., Loreau, M., and Haegeman, B. (2019). The inherent multidimensionality of temporal

variability: how common and rare species shape stability patterns. *Ecol. Lett.* 22, 1557–1567. doi:10.1111/ele.13345.

Baert, J. M., De Laender, F., Sabbe, K., and Janssen, C. R. (2016). Biodiversity increases functional and compositional resistance, but decreases resilience in phytoplankton communities. *Ecology* 97, 3433–3440. doi:10.1002/ecy.1601.

Carrara, F., Altermatt, F., Rodriguez-Iturbe, I., and Rinaldo, A. (2012). Dendritic connectivity controls biodiversity patterns in experimental metacommunities. *Proc. Natl. Acad. Sci. U. S. A.* 109, 5761–5766. doi:10.1073/pnas.1119651109.

Carraro, L., Bertuzzo, E., Fronhofer, E. A., Furrer, R., Gounand, I., Rinaldo, A., et al. (2020). Generation and application of river network analogues for use in ecology and evolution. *Ecol. Evol.* 10, 7537–7550. doi:10.1002/ece3.6479.

Chase, J. M., Blowes, S. A., Knight, T. M., Gerstner, K., and May, F. (2020). Ecosystem decay exacerbates biodiversity loss with habitat loss. *Nature* 584, 238–243. doi:10.1038/s41586-020-2531-2.

Chesson, P. (2000). General theory of competitive coexistence in spatially-varying environments. *Theor. Popul. Biol.* 58, 211–37. doi:10.1006/tpbi.2000.1486.

Clark, A. T., Arnoldi, J. F., Zelnik, Y. R., Barabas, G., Hodapp, D., Karakoç, C., et al. (2021). General statistical scaling laws for stability in ecological systems. *Ecol. Lett.* 24, 1474–1486. doi:10.1111/ele.13760.

Cochran, W. G. (1977). *Sampling Techniques*. 3rd editio. New York: John Wiley & Sons.

De Raedt, J., Baert, J. M., Janssen, C. R., and De Laender, F. (2019). Stressor fluxes alter the relationship between beta-diversity and regional productivity. *Oikos* 128, 1015–1026. doi:10.1111/oik.05191.

Doak, D. F., Bigger, D., Harding, E. K., Marvier, M. A., O'Malley, R. E., and Thomson, D. (1998). The statistical inevitability of stability-diversity relationships in community ecology. *Am. Nat.* 151, 264–276. doi:10.1086/286117.

Domínguez-García, V., Dakos, V., and Kéfi, S. (2019). Unveiling dimensions of stability in complex ecological networks. *Proc. Natl. Acad. Sci. U. S. A.* 116, 25714–25720. doi:10.1073/pnas.1904470116.

Donohue, I., Petchey, O. L., Montoya, J. M., Jackson, A. L., Mcnally, L., Viana, M., et al. (2013). On the dimensionality of ecological stability. *Ecol. Lett.* 16, 421–429. doi:10.1111/ele.12086.

Downing, A. L., Brown, B. L., and Leibold, M. A. (2014). Multiple diversity-stability mechanisms enhance population and community stability in aquatic food webs. *Ecology* 95, 173–184. doi:10.1890/12-1406.1.

Efron, B., and Tibshirani, R. (1985). The Bootstrap Method for Assessing Statistical Accuracy. *Behaviormetrika* 12, 1–35. doi:10.2333/bhmk.12.17_1.

Fagan, W. F. (2002). Connectivity, fragmentation, and extinction risk in dendritic metapopulations. *Ecology* 83, 3243–3249. doi:10.1890/0012-9658(2002)083[3243:CFAERI]2.0.CO;2.

Ferger, W. F. (1931). The Nature and Use of the Harmonic Mean. *J. Am. Stat. Assoc.* 26, 36–40. doi:10.1080/01621459.1931.10503148.

Flöder, S., and Hillebrand, H. (2012). Species traits and species diversity affect community stability in a multiple stressor framework. *Aquat. Biol.* 17, 197–209. doi:10.3354/ab00479.

Gatz, D. F., and Smith, L. (1995). The standard error of a weighted mean concentration-I. Bootstrapping vs other methods. *Atmos. Environ.* 29, 1185–1193. doi:10.1016/1352-2310(94)00210-C.

Gravel, D., Massol, F., and Leibold, M. A. (2016). Stability and complexity in model meta-ecosystems. *Nat. Commun.* 7, 12457. doi:10.1038/ncomms12457.

Greig, H. S., McHugh, P. A., Thompson, R. M., Warburton, H. J., and McIntosh, A. R. (2022). Habitat size influences community stability. *Ecology* 103, e03545. doi:10.1002/ecy.3545.

Grimm, V., and Wissel, C. (1997). Babel, or the ecological stability discussions: An inventory and analysis of terminology and a guide for avoiding confusion. *Oecologia* 109, 323–334. doi:10.1007/s004420050090.

Gross, K., Cardinale, B. J., Fox, J. W., Gonzalez, A., Loreau, M., Wayne Polley, H., et al. (2014). Species richness and the temporal stability of biomass production: A new analysis of recent biodiversity experiments. *Am. Nat.* 183, 1–12. doi:10.1086/673915.

Haegeman, B., Arnoldi, J.-F., Wang, S., de Mazancourt, C., Montoya, J., and Loreau, M. (2016). Resilience, Invariability, and Ecological Stability across Levels of Organization. *bioRxiv*, 085852. doi:10.1101/085852.

Hesterberg, T. (2011). Bootstrap. *Wiley Interdiscip. Rev. Comput. Stat.* 3, 497–526. doi:10.1002/wics.182.

Hillebrand, H., Langenheder, S., Lebret, K., Lindström, E., Östman, Ö., and Striebel, M. (2018). Decomposing multiple dimensions of stability in global change experiments. *Ecol. Lett.* 21, 21–30. doi:10.1111/ele.12867.

Isbell, F., Craven, D., Connolly, J., Loreau, M., Schmid, B., Beierkuhnlein, C., et al. (2015). Biodiversity increases the resistance of ecosystem productivity to climate extremes. *Nature* 526, 574–577. doi:10.1038/nature15374.

IUPAC (2019). *The IUPAC Compendium of Chemical Terminology - The Gold Book*. 2nd ed. , ed. V. Gold Research Triangle Park, NC: International Union of Pure and Applied Chemistry (IUPAC) doi:10.1351/goldbook.

Ives, A. R., and Carpenter, S. R. (2007). Stability and diversity of ecosystems. *Science (80-. ).* 317, 58–62. doi:10.1126/science.1133258.

Ives, A. R., Klug, J. L., and Gross, K. (2000). Stability and species richness in complex communities.

*Ecol. Lett.* 3, 399–411. doi:10.1046/j.1461-0248.2000.00144.x.

Jarillo, J., Sæther, B.-E., Engen, S., and Cao-García, F. J. (2020). Spatial Scales of Population Synchrony in Predator-Prey Systems. *Am. Nat.* 195, 216–230. doi:10.1086/706913.

Jarillo, J., Sæther, B.-E., Engen, S., and Cao, F. J. (2018). Spatial scales of population synchrony of two competing species: effects of harvesting and strength of competition. *Oikos* 127, 1459–1470. doi:10.1111/oik.05069.

Karakoç, C., Clark, A. T., and Chatzinotas, A. (2020). Diversity and coexistence are influenced by time-dependent species interactions in a predator–prey system. *Ecol. Lett.* 23, 983–993. doi:10.1111/ele.13500.

Kéfi, S., Domínguez-García, V., Donohue, I., Fontaine, C., Thébault, E., and Dakos, V. (2019). Advancing our understanding of ecological stability. *Ecol. Lett.* 22, 1349–1356. doi:10.1111/ele.13340.

Lande, R., Engen, S., and Sæther, B.-E. (1999). Spatial Scale of Population Synchrony: Environmental Correlation versus Dispersal and Density Regulation. *Am. Nat.* 154, 271–281. doi:10.1086/303240.

Leibold, M. A., Holyoak, M., Mouquet, N., Amarasekare, P., Chase, J. M., Hoopes, M. F., et al. (2004). The metacommunity concept: A framework for multi-scale community ecology. *Ecol. Lett.* 7, 601–613. doi:10.1111/j.1461-0248.2004.00608.x.

Lemoine, N. P. (2020). Unifying ecosystem responses to disturbance into a single statistical framework. *Oikos*, 1–14. doi:10.1111/oik.07752.

Levin, S. A. (1992). The Problem of Pattern and Scale in Ecology: The Robert H. MacArthur Award Lecture. *Ecology* 73, 1943–1967. doi:10.2307/1941447.

Limberger, R., Pitt, A., Hahn, M. W., and Wickham, S. A. (2019). Spatial insurance in multi-trophic metacommunities. *Ecol. Lett.* 22, 1828–1837. doi:10.1111/ele.13365.

Liu, J., Soininen, J., Han, B. P., and Declerck, S. A. J. (2013). Effects of connectivity, dispersal directionality and functional traits on the metacommunity structure of river benthic diatoms. *J. Biogeogr.* 40, 2238–2248. doi:10.1111/jbi.12160.

Loreau, M., and De Mazancourt, C. (2008). Species synchrony and its drivers: Neutral and nonneutral community dynamics in fluctuating environments. *Am. Nat.* 172, 48–66. doi:10.1086/589746.

May, R. M. (1972). Will a Large Complex System be Stable? *Nature* 238, 413–414. doi:10.1038/238413a0.

McCann, K. S. (2000). The diversity-stability debate. *Nature* 405, 228–33. doi:10.1038/35012234.

McCluney, K. E., Poff, N. L., Palmer, M. A., Thorp, J. H., Poole, G. C., Williams, B. S., et al. (2014). Riverine macrosystems ecology: Sensitivity, resistance, and resilience of whole river basins with human alterations. *Front. Ecol. Environ.* 12, 48–58. doi:10.1890/120367.

Moran, P. A. P. (1953). The statistical analysis of the Canadian Lynx cycle. II. Synchronization and Meteorology. *Aust. J. Zool.* 1, 291–298. doi:10.1071/ZO9530291.

Mougi, A., and Kondoh, M. (2012). Diversity of interaction types and ecological community stability. *Science (80-. ).* 337, 349–351. doi:10.1126/science.1220529.

Mougi, A., and Kondoh, M. (2016). Food-web complexity, meta-community complexity and community stability. *Sci. Rep.* 6, 1–5. doi:10.1038/srep24478.

Neubert, M. G., and Caswell, H. (1997). Alternatives to resilience for measuring the responses of ecological systems to perturbations. *Ecology* 78, 653–665. doi:10.1890/0012-9658(1997)078[0653:ATRFMT]2.0.CO;2.

Pennekamp, F., Pontarp, M., Tabi, A., Altermatt, F., Alther, R., Choffat, Y., et al. (2018). Biodiversity increases and decreases ecosystem stability. *Nature* 563, 109–112. doi:10.1038/s41586-018-0627-8.

Peterson, E. E., Ver Hoef, J. M., Isaak, D. J., Falke, J. A., Fortin, M. J., Jordan, C. E., et al. (2013). Modelling dendritic ecological networks in space: An integrated network perspective. *Ecol. Lett.* 16, 707–719. doi:10.1111/ele.12084.

Pimm, S. L. (1984). The complexity and stability of ecosystems. *Nature* 307, 321–326. doi:10.1038/307321a0.

Plitzko, S. J., and Drossel, B. (2015). The effect of dispersal between patches on the stability of large trophic food webs. *Theor. Ecol.* 8, 233–244. doi:10.1007/s12080-014-0247-3.

Python Core Team (2019). Python: A dynamic, open source programming language. Available at: https://www.python.org/.

R Core Team (2020). R: A Language and Environment for Statistical Computing. Available at: https://www.r-project.org/.

Radchuk, V., Laender, F. De, Cabral, J. S., Boulangeat, I., Crawford, M., Bohn, F., et al. (2019). The dimensionality of stability depends on disturbance type. *Ecol. Lett.* 22, 674–684. doi:10.1111/ele.13226.

Rooney, N., McCann, K., Gellner, G., and Moore, J. C. (2006). Structural asymmetry and the stability of diverse food webs. *Nature* 442, 265–269. doi:10.1038/nature04887.

Saade, C., Kéfi, S., Gougat-Barbera, C., Rosenbaum, B., and Fronhofer, E. A. (2020). Spatial distribution of local patch extinctions drives recovery dynamics in metacommunities. *bioRxiv*, 2020.12.03.409524. Available at: http://biorxiv.org/content/early/2020/12/04/2020.12.03.409524.abstract.

Saeedian, M., Pigani, E., Maritan, A., Suweis, S., and Azaele, S. (2021). Effect of delay on the emergent stability patterns in Generalized Lotka-Volterra ecological dynamics. 1–18. Available at: http://arxiv.org/abs/2110.11914.

Stevens, S. S. (1955). On the Averaging of Data. *Science (80-. ).* 121, 113–116.

doi:10.1126/science.121.3135.113.

Supp, S. R., and Ernest, S. K. M. (2014). Species-level and community-level responses to disturbance: A cross-community analysis. *Ecology* 95, 1717–1723. doi:10.1890/13-2250.1.

Thébault, E., and Loreau, M. (2005). Trophic interactions and the relationship between species diversity and ecosystem stability. *Am. Nat.* 166, E95-114. doi:10.1086/444403.

Thibaut, L. M., and Connolly, S. R. (2013). Understanding diversity-stability relationships: Towards a unified model of portfolio effects. *Ecol. Lett.* 16, 140–150. doi:10.1111/ele.12019.

Tilman, D., Reich, P. B., and Knops, J. M. H. (2006). Biodiversity and ecosystem stability in a decade-long grassland experiment. *Nature* 441, 629–632. doi:10.1038/nature04742.

Wang, S., Lamy, T., Hallett, L. M., and Loreau, M. (2019). Stability and synchrony across ecological hierarchies in heterogeneous metacommunities: linking theory to data. *Ecography (Cop.).* 42, 1200–1211. doi:10.1111/ecog.04290.

Wang, S., and Loreau, M. (2014). Ecosystem stability in space: α, β and γ variability. *Ecol. Lett.* 17, 891–901. doi:10.1111/ele.12292.

Wang, S., Loreau, M., Arnoldi, J. F., Fang, J., Rahman, K. A., Tao, S., et al. (2017). An invariability-area relationship sheds new light on the spatial scaling of ecological stability. *Nat. Commun.* 8, 1–8. doi:10.1038/ncomms15211.

Yachi, S., and Loreau, M. (1999). Biodiversity and ecosystem productivity in a fluctuating environment: The insurance hypothesis. *Proc. Natl. Acad. Sci. U. S. A.* 96, 1463–1468. doi:10.1073/pnas.96.4.1463.

Yang, Q., Fowler, M. S., Jackson, A. L., and Donohue, I. (2019). The predictability of ecological stability in a noisy world. *Nat. Ecol. Evol.* 3, 251–259. doi:10.1038/s41559-018-0794-x.

Yodzis, P. (1981). The stability of real ecosystems. *Nature* 289, 674–676. doi:10.1038/289674a0.

## *Supplementary Material*

**Appendix A: Standard errors of stability components**

### A.1. Growth rate

Section 3 of the main text shows that local population biomasses and local population-level growth rates suffice to estimate the growth rate of the spatial networks of communities (Eqs. (7-9)). The network-level growth rate $R_{net}$ is the arithmetic mean of node-level growth, weighted by the node biomasses. However, in real networks, measuring the growth rate at each of the nodes of the network might be not feasible. The level of uncertainty that we will have in our regional or community estimate $\widetilde{R_{net}}$ coming from an incomplete number of measurements will be given by the standard error on this estimate. The standard error of a weighted arithmetic mean (here defined as half the width of the 95% confidence interval of the weighted mean from a sampling of size $\tilde{n}$) follows the expression obtained by Cochran (1977), and validated with bootstrapping techniques (Gatz and Smith, 1995),

$$\text{SE}(\widetilde{R_{net}}) = t_{\tilde{n}-1} \sqrt{\frac{\tilde{n}}{(\tilde{n}-1)(\sum_i N_i)^2}} \times$$

$$\times \sqrt{\left[\sum_{i=1}^{\tilde{n}} \left(N_i R_i - M_N \widetilde{R_{net}}\right)^2 - 2\widetilde{R_{net}} \sum_{i=1}^{\tilde{n}} (N_i - M_N)\left(N_i R_i - M_N \widetilde{R_{net}}\right) + \left(\widetilde{R_{net}}\right)^2 \sum_{i=1}^{\tilde{n}} (N_i - M_N)^2\right]}, \text{(A1)}$$

where $M_N$ is the unweighted arithmetic sample mean of the biomasses, $M_N \equiv \frac{1}{\tilde{n}}\sum_{i=1}^{\tilde{n}} N_i$ (see Table 1 of the main text); $\widetilde{R_{net}} \equiv \sum_{i=1}^{\tilde{n}} N_i \rho_i / \sum_{i=1}^{n} N_i$ is the estimate of network growth rate, obtained as the weighted arithmetic mean of growth rates of the sampled species or locations, with weights equal to the local or species biomasses $N$; and $\tilde{n}$ is the number of sampled locations or species. $t_{\tilde{n}-1}$ is the Student t-distribution with $\tilde{n} - 1$ degrees of freedom corresponding to the 95% confidence interval. Its value for large enough sampling sizes ($\tilde{n} \to \infty$) is approximately 1.96. Using the general expression for the covariance of the product of random variables (Bohrnstedt and Goldberger, 1969), we can write the covariance of $N$ with the product $N \cdot R$ as $\text{cov}(N, N\,R) \equiv \frac{1}{\tilde{n}}\sum_{i=1}^{\tilde{n}}(N_i - M_N)(N_i R_i - M_{N\times R}) = M_N \text{cov}(N, R) + M_R \text{var}(N)$ (where "cov" denotes the sampling covariance , $\text{cov}(N, R) \equiv M_{N\cdot R} - M_N M_R$, and "var" the sampling variance, $\text{var}(N) \equiv S_N^2 \equiv M_{N^2} - M_N^2$). Then, this general expression can be further reduced to

$$\text{SE}(\widetilde{R_{net}}) = t_{\tilde{n}-1} \frac{S_R}{\sqrt{\tilde{n}-1}} \sqrt{1 + \left(1 - C_{N,R}^2\right)\left(\frac{S_N}{M_N}\right)^2 + C_{N,R}^2 \left(\frac{S_N}{M_N}\right)^4}, \tag{A2}$$

where $S_R = \sqrt{\text{var}(R)}$ is the standard deviation of the sampled values of growth rate at the nodes of the network, $S_N/M_N$ is the coefficient of variation of the node biomasses (the ratio of the sample standard deviation $S_N$ and the sample mean $M_N$ of the node biomasses), and $C_{N,R} \equiv \text{cov}(N,R)/(S_N S_R)$ is the correlation of the nodes biomasses and growth rates (see Table 1 of the main text). Note that for the special case of no variation in the local or species biomasses, the standard error of the unweighted arithmetic mean is recovered, $SE(M_R) = t_{\tilde{n}-1} \frac{S_R}{\sqrt{\tilde{n}-1}}$. The standard error estimated with Eq. (A2) equals the standard error that can be obtained numerically with bootstrapping techniques (Efron and Tibshirani, 1985; Hesterberg, 2011), although in our simulations it produces a slight overestimation (Figs. S7-S8). It is important to note that this analytical expression is valid for any possible community or spatial network, regardless of its complexity. We can further normalize by the network growth rate

$$\widetilde{R_{net}} = \frac{\sum_i N_i \rho_i}{\sum_i N_i} = M_R \left(1 + \frac{S_N}{M_N} \frac{S_R}{M_R} C_{N,R}\right)$$

to obtain an expression for the relative standard error

$$\frac{SE(\widetilde{R_{net}})}{\widetilde{R_{net}}} = \frac{t_{n-1}}{\sqrt{n-1}} \frac{\frac{S_R}{M_R}}{1 + \frac{S_N}{M_N} \frac{S_R}{M_R} C_{N,R}} \sqrt{1 + \left(1 - C_{N,R}^2\right) \left(\frac{S_N}{M_N}\right)^2 + C_{N,R}^2 \left(\frac{S_N}{M_N}\right)^4}. \quad (A3)$$

Using Eq. (A3), we can estimate the required sample size needed to control the standard error of the growth rate estimate of a sampled network based on the coefficients of variation and on the correlation of the node biomasses and growth rates (Fig. 3 of the main text, Fig. S6). This required sample size is mainly determined by the coefficient of variation of the node growth rates (Fig. 3), having the coefficient of variation of the node biomasses and the correlation between the biomass and the growth rate smaller effects except for the case of high anticorrelation (Fig. S6).

## A.2. Resistance

Network resistance has been obtained as the weighted harmonic mean of local or species resistances (Eq. (12) of the main text, and Appendix B). Hence, it is easy to check that the network estimate of the inverse of resistance, $\Omega^{-1}$, would simply be given as the weighted arithmetic mean of the individual values. The standard error of $\Omega^{-1}$ would then be equivalent to that expressed in Eq. (A2), changing $R$ by $\Omega^{-1}$,

$$SE\left((\widetilde{\Omega^{-1}})_{net}\right) = t_{n-1} \frac{S_{\Omega^{-1}}}{\sqrt{n-1}} \sqrt{1 + \left(1 - C_{N,\Omega^{-1}}^2\right) \left(\frac{S_N}{M_N}\right)^2 + C_{N,\Omega^{-1}}^2 \left(\frac{S_N}{M_N}\right)^4}. \quad (A4)$$

Moreover, the standard error of the unweighted harmonic mean of any random variable $X$ is just equal to the standard error of the arithmetic mean of $X^{-1}$, multiplied by the squared harmonic mean of $X$ (Norris, 1940). Assuming this approximately holds also for the weighted case, we would obtain that the standard error of the network resistance is

$$SE(\widetilde{\Omega_{net}}) = \left(\widetilde{\Omega_{net}}\right)^2 SE\left((\widetilde{\Omega^{-1}})_{net}\right)$$

$$= \left(\widetilde{\Omega_{net}}\right)^2 t_{n-1} \frac{S_{\Omega^{-1}}}{\sqrt{n-1}} \sqrt{1 + \left(1 - C_{N,\Omega^{-1}}^2\right) \left(\frac{S_N}{M_N}\right)^2 + C_{N,\Omega^{-1}}^2 \left(\frac{S_N}{M_N}\right)^4}$$

$$= \frac{t_{n-1}}{\sqrt{n-1}} \frac{S_{\Omega^{-1}}}{\left(M_{\Omega^{-1}}\right)^2} \left(\frac{1}{1 + \frac{S_N}{M_N}\frac{S_{\Omega^{-1}}}{M_{\Omega^{-1}}}C_{N,\Omega^{-1}}}\right)^2 \sqrt{1 + \left(1 - C_{N,\Omega^{-1}}^2\right)\left(\frac{S_N}{M_N}\right)^2 + C_{N,\Omega^{-1}}^2\left(\frac{S_N}{M_N}\right)^4} \quad \text{(A5)}$$

being

$$\widetilde{\Omega_{net}} \equiv \frac{\sum_i N_i}{\sum_i N_i\,\Omega_i^{-1}} = \frac{1}{M_{\Omega^{-1}}}\frac{1}{1 + \frac{S_N}{M_N}\frac{S_{\Omega^{-1}}}{M_{\Omega^{-1}}}C_{N,\Omega^{-1}}}$$

the harmonic mean of the resistances of the sampled nodes weighted by the biomasses. Again, this analytical expression for the standard error of the resistance produces results compatible with those obtained numerically with a bootstrap (Figs. S7-S8). Moreover, normalizing by the network resistance $\widetilde{\Omega_{net}}$, we recover an expression analogous to Eq. (A3),

$$\frac{\mathrm{SE}(\widetilde{\Omega_{net}})}{\widetilde{\Omega_{net}}} = \frac{t_{n-1}}{\sqrt{n-1}}\frac{\frac{S_{\Omega^{-1}}}{M_{\Omega^{-1}}}}{1 + \frac{S_N}{M_N}\frac{S_{\Omega^{-1}}}{M_{\Omega^{-1}}}C_{N,\Omega^{-1}}}\sqrt{1 + \left(1 - C_{N,\Omega^{-1}}^2\right)\left(\frac{S_N}{M_N}\right)^2 + C_{N,\Omega^{-1}}^2\left(\frac{S_N}{M_N}\right)^4} \quad . \quad \text{(A6)}$$

From Eq. (A6) it is possible to estimate the sample size required to control the error of the resistance estimated of a sampled ecological network. Analogously than for the growth rate, the higher element which will mainly determine this sample size is the coefficient of variation of the reciprocal of the node resistance, $S_{\Omega^{-1}}/M_{\Omega^{-1}}$, except for the case of high anticorrelation between biomass and the reciprocal of resistance.

### A.3. Initial resilience

With respect to the initial resilience, in the main text we have seen that it is just the product of the growth rate and the resistance, $\rho = R\,\Omega$. Hence, we can write the network estimate of initial resilience as the product of the estimates of the growth rates and resistances, $\widetilde{\rho_{net}} = \widetilde{R_{net}}\,\widetilde{\Omega_{net}}$. And assuming no correlations between the growth rates and the resistances, the standard error of the estimated network initial resilience would be $\mathrm{SE}(\widetilde{\rho_{net}}) = \widetilde{\Omega_{net}}\,\mathrm{SE}(\widetilde{R_{net}}) + \widetilde{R_{net}}\,\mathrm{SE}(\widetilde{\Omega_{net}})$. Hence, with the estimations of the network growth rate after a perturbation and resistance we can obtain the estimation (and the standard error) of the network initial resilience.

### A.4. Invariability

For the case of invariability, the results are more complicated. For the particular case of perfectly synchronous dynamics, the network invariability is the square of the weighted harmonic mean of the square roots of the invariabilities (Eq. (20) main text). The standard error of the harmonic mean of $\sqrt{I_i^s}$ would then given by Eq. (A5), replacing $\Omega$ by $\sqrt{I^s}$, and the biomasses $N$ by their steady states $N^*$. Regarding the standard error of the square of such mean, the results can be easily computed via error propagation, achieving the estimate

$$\mathrm{SE}(\widetilde{I_{net}^s}) = 2\,t_{n-1}\,(\widetilde{I_{net}^s})^{\frac{3}{2}}\,\frac{S_{\sqrt{I}^{-1}}}{\sqrt{n-1}}\sqrt{1 + \left(1 - C_{N,\sqrt{I}^{-1}}^2\right)\left(\frac{S_{N^*}}{M_{N^*}}\right)^2 + \left(\frac{S_{N^*}}{M_{N^*}}\right)^4 C_{N,\sqrt{I}^{-1}}^2} \quad . \quad \text{(A7)}$$

However, the more general cases of asynchronous or partially asynchronous dynamics are more involved, since for them the invariability is not a mean of individual invariabilities, but depend on the size of the network. Hence, to assess the uncertainty of the network invariability, we would generally not be able to employ any of these analytical expression, and we should compute it numerically with bootstrapping techniques (Efron and Tibshirani, 1985; Hesterberg, 2011).

## Appendix B: Scaling of resistance

We define the resistance as the inverse of the relative change of the biomass caused by a perturbation at time $t_0$, (Eq. (2)),

$$\Omega = \frac{N(t_0)}{N(t_0) - N(t_0 + \delta t)} \ .$$ (B1)

Hence, the resistance of a species $i$ located at $x$ would be:

$$\Omega_{x,i} \equiv \frac{N_{x,i}(t_0)}{N_{x,i}(t_0) - N_{x,i}(t_0 + \delta t)} \ .$$ (B2)

Regional resistance of a species $i$ is defined as the resistance of the total biomass of such species across all locations. Then, applying the definition in Eq. (B1) to that regional biomass of the species $i$, $N_i \equiv \sum_x N_{x,i}$, it is easy to see that

$$\Omega_i^{(\mathcal{R})} \equiv \frac{N_i(t_0)}{N_i(t_0) - N_i(t_0 + \delta t)} = \left[ \frac{\sum_x N_{x,i}(t_0) \frac{1}{\Omega_{x,i}}}{\sum_x N_{x,i}(t_0)} \right]^{-1} ,$$ (B3)

which is the definition of the harmonic mean of local species resistances $\Omega_{x,i}$, weighted by the local species biomasses before the perturbation. It can also be seen that the local community resistance (which is the resistance of the sum of biomasses across all species at a specific location, $N_x \equiv \sum_i N_{x,i}$) is also the weighted harmonic mean of species resistances,

$$\Omega_x^{(\mathcal{C})} \equiv \frac{N_x(t_0)}{N_x(t_0) - N_x(t_0 + \delta t)} = \left[ \frac{\sum_i N_{x,i}(t_0) \frac{1}{\Omega_{x,i}}}{\sum_i N_{x,i}(t_0)} \right]^{-1} \ .$$ (B4)

Finally, the regional community resistance is the weighted harmonic mean of local community resistances, or the weighted harmonic mean of regional species resistances,

$$\Omega^{(\mathcal{RC})} \equiv \frac{N_T(t_0)}{N_T(t_0) - N_T(t_0 + \delta t)} = \left[ \frac{\sum_x \sum_i N_{x,i}(t_0) \frac{1}{\Omega_{x,i}}}{\sum_x \sum_i N_{x,i}(t_0)} \right]^{-1} = \left[ \frac{\sum_x N_x(t_0) \frac{1}{\Omega_x^{(\mathcal{C})}}}{\sum_x N_x(t_0)} \right]^{-1} = \left[ \frac{\sum_i N_i(t_0) \frac{1}{\Omega_i^{(\mathcal{R})}}}{\sum_i N_i(t_0)} \right]^{-1} \ .$$ (B5)

In general, the resistance of any ecological or spatial network can be rewritten as

$$\Omega_{net} = \frac{1}{\mu_{\Omega^{-1}}} \frac{1}{1 + \frac{\sigma_N}{\mu_N} \frac{\sigma_{\Omega^{-1}}}{\mu_{\Omega^{-1}}} c_{N,\Omega^{-1}}}, \tag{B6}$$

where $\Omega_{net}$ represents the resistance of any ecological or spatial network, $\mu_{\Omega^{-1}}$ is the unweighted population mean of the node estimates of resistance (so $1/\mu_{\Omega^{-1}}$ represents the unweighted population harmonic mean of node resistances); $\sigma_{\Omega^{-1}} = \sqrt{\mathrm{var}(\Omega^{-1})}$ the standard deviation of the reciprocals of the node resistances, and $c_{N,\Omega^{-1}} = \mathrm{cov}(N,\Omega^{-1})/(\sigma_N \sigma_{\Omega^{-1}})$ the normalized correlation (see Table 1 of the main text). This equation also indicates that correlations between the node biomasses and the inverse of the resistance can make the network resistance $\Omega_{net}$ greater or smaller than the unweighted harmonic mean of resistances.

Eq. (B6) can also be applied to estimate the network resistance $\Omega_{net}$ when only a limited number of nodes have been sampled from the network, so the sample estimate of resistance would be

$$\widetilde{\Omega_{net}} = \frac{1}{M_{\Omega^{-1}}} \frac{1}{1 + \frac{S_N}{M_N} \frac{S_{\Omega^{-1}}}{M_{\Omega^{-1}}} C_{N,\Omega^{-1}}}, \tag{B7}$$

where again $M$ represent sample means, $S$ sample standard deviations, and $C$ sample correlations (see Table 1).

Even though the resistance of a network $\Omega_{net}$ is the weighted harmonic mean of resistances, the network estimate of its inverse $(\Omega^{-1})_{net}$ is the weighted arithmetic mean of the sub-units estimates of $\Omega^{-1}$. This result again allows us to estimate its standard error arising from incomplete but representative network sampling (Appendix A, Eq. (A5)), while the relative standard error follows a formula analogous to that of the growth (Appendix A, Eq. (A6)). Hence, the relative uncertainty of the network resistance will be dominated by the number of samples from the network, and by the variance of the inverse of resistances.

## Appendix C: Relation Between Weighted Harmonic and Arithmetic Mean of Resistances

The regional or community resistance has been obtained to be the harmonic mean of the local or population resistances, weighted by the biomass. (Eq. (13) of the main text). Here, we want to prove that such weighted harmonic mean is upper-bounded by the weighted arithmetic mean of resistances, *i.e.*,

$$\frac{1}{\sum_a n_a \frac{1}{\Omega_a}} \leq \sum_a n_a \, \Omega_a \; . \tag{C1}$$

We denote $\nu_a(t_0) \equiv N_a(t_0)/\sum_b N_b(t_0)$ the proportion of the biomass of the network at the node $a$. Such node could represent either a location or a species. Prove the relation (C1) is equivalent to prove that

$$\Psi \equiv \left( \sum_a \nu_a \, \Omega_a \right) \times \left( \sum_b \nu_b \frac{1}{\Omega_b} \right) \geq 1 \,. \tag{C2}$$

Let's start to simplify the expression in Eq. (C2)

$$\Psi \equiv \left( \sum_a \nu_a \, \Omega_a \right) \times \left( \sum_b \nu_b \frac{1}{\Omega_b} \right) = \sum_a \sum_b \nu_a \, \nu_b \frac{\Omega_a}{\Omega_b}$$

$$= \sum_a \sum_b \nu_a \, \nu_b - \sum_a \sum_b \nu_a \, \nu_b + \sum_a \sum_b \nu_a \, \nu_b \frac{\Omega_a}{\Omega_b} = \left( \sum_a \nu_a \right)^2 + \sum_a \sum_b \nu_a \, \nu_b \left( \frac{\Omega_a}{\Omega_b} - 1 \right).$$

From the definition of $\nu_a$, $\sum_a \nu_a = 1$ (the sum of proportions over all nodes of the networks is 1). Hence

$$\Psi = 1 + \sum_a \sum_b \nu_a \, \nu_b \left( \frac{\Omega_a}{\Omega_b} - 1 \right)$$

$$= 1 + \sum_a \nu_a^2 \, (1-1) + \sum_a \sum_{b>a} \nu_a \, \nu_b \left( \frac{\Omega_a}{\Omega_b} - 1 \right) + \sum_a \sum_{b<a} \nu_a \, \nu_b \left( \frac{\Omega_a}{\Omega_b} - 1 \right)$$

$$= 1 + \sum_a \sum_{b>a} \nu_a \, \nu_b \left( \frac{\Omega_a}{\Omega_b} - 1 \right) + \sum_{a'} \sum_{b'>a'} \nu_{a'} \, \nu_{b'} \left( \frac{\Omega_{b'}}{\Omega_{a'}} - 1 \right)$$

$$= 1 + \sum_a \sum_{b>a} \nu_a \, \nu_b \left( \frac{\Omega_a}{\Omega_b} + \frac{\Omega_b}{\Omega_a} - 2 \right).$$

But also

$$\frac{\Omega_a}{\Omega_b} + \frac{\Omega_b}{\Omega_a} - 2 = \frac{\Omega_a^2 + \Omega_b^2 - 2 \, \Omega_a \, \Omega_b}{\Omega_a \, \Omega_b} = \frac{(\Omega_a - \Omega_b)^2}{\Omega_a \, \Omega_b},$$

so

$$\Psi = 1 + \sum_a \sum_{b>a} \nu_a \, \nu_b \frac{(\Omega_a - \Omega_b)^2}{\Omega_a \, \Omega_b} \,. \tag{C3}$$

All the terms $\nu_a$, $\Omega_a$, and $(\Omega_a - \Omega_b)^2$ are non-negative. Hence, the second term in Eq. (C3) is also non-negative, and in consequence

$$\Psi \geq 1 \,, \tag{C4}$$

where $\Psi = 1$ only if $\Omega_a = \Omega_b$ for each pair of nodes of the network with non-zero biomass.

Eq. (C4) proves that Eq (C2) is true, so it proves that the weighted harmonic mean of resistances is upper-bounded by the weighted arithmetic mean of resistances (Eq. (C1)). And both means are equal only if there is no variation in the individual resistances of the nodes with non-zero biomass.

This result implies that the network resistance estimate, equal to the weighted harmonic mean of the node resistance estimates, is upper bounded by the weighted arithmetic mean of resistances, being more affected by low-resistant nodes than the arithmetic mean is.

**Appendix D: Scaling of invariability**

We define invariability $I$ as the ratio of the square temporal mean of the biomass and its temporal variance:

$$I \equiv \frac{\left[\text{mean}_t\big(N(t)\big)\right]^2}{\text{var}_t\big(N(t)\big)} \ .$$ 

(D1)

In general meta-community models, we define the local species invariability as the invariability of the local species population,

$$I_{x,i} \equiv \frac{\left[\text{mean}_t\left(N_{x,i}(t)\right)\right]^2}{\text{var}_t\left(N_{x,i}(t)\right)} \ .$$ 

(D2)

The regional invariability is defined as the invariability of the sum of local biomasses of a specific species across locations, $N_i \equiv \sum_x N_{x,i}$. If the biomass of species $i$ at location $x$ is fluctuating around a steady state, with a temporal mean $N_{x,i}^* \equiv \text{mean}_t\left(N_{x,i}(t)\right)$ and a temporal variance $\left[\text{std}_t\left(N_{x,i}\right)\right]^2 \equiv \text{var}_t\left(N_{x,i}(t)\right)$, from Eq. (D1) we can express the temporal standard deviation as

$$\text{std}_t\left(N_{x,i}\right) \equiv \frac{N_{x,i}^*}{\sqrt{I_{x,i}}} \ .$$ 

(D3)

We define the regional invariability as the invariability of the total biomass of the species across the multiple locations of the spatial network,

$$I_i^{(\mathcal{R})} \equiv \frac{\left[\text{mean}_t\big(\sum_x N_{x,i}(t)\big)\right]^2}{\text{var}_t\big(\sum_x N_{x,i}(t)\big)} \ .$$ 

(D4)

The numerator in Eq. (D4) is simply the square of the sum of the temporal means of the local biomasses,

$$\left[\text{mean}_t\left(\sum_x N_{x,i}(t)\right)\right]^2 = \left[\sum_x \text{mean}_t\left(N_{x,i}(t)\right)\right]^2 = \left(\sum_x N_{x,i}^*\right)^2 \ .$$ 

(D5)

The denominator of Eq. (D4) can be decomposed as the sum of temporal covariances for each pair of locations in the spatial network,

$$\text{var}_t\left(\sum_x N_{x,i}(t)\right) = \sum_x \sum_y \text{cov}_t\left(N_{x,i}, N_{y,i}\right) \equiv \sum_x \sum_y \text{std}_t(N_{x,i})\ \text{std}_t(N_{y,i})\ \text{corr}_t\left(N_{x,i}, N_{y,i}\right), \quad \text{(D6)}$$

where $\text{cov}_t\left(N_{x,i}, N_{y,i}\right)$ is the temporal covariance of the local biomasses, and $\text{corr}_t\left(N_{x,i}, N_{y,i}\right)$ is the temporal correlation of local biomasses. Replacing Eqs. (D5) and (D6) in Eq. (D4), and expressing the temporal standard deviations as shown in Eq. (D3), we finally get

$$I_i^{(\mathcal{R})} = \frac{\left(\sum_x N_{x,i}^*\right)^2}{\sum_x \sum_y N_{x,i}^* N_{y,i}^* \text{corr}_t\left(N_{x,i}, N_{y,i}\right) \frac{1}{\sqrt{I_{x,i}}} \frac{1}{\sqrt{I_{y,i}}}}, \quad \text{(D7)}$$

Eq. (D7) gives the general expression of the regional invariability as a function of local invariabilities $I_{x,i}$, but it can be informative to study two limiting cases. If the population dynamics were uncorrelated for different locations ($\text{corr}_t\left(N_{x,i}, N_{y,i}\right) = \delta_{x,y}$), this regional invariability (Eq. (D7)) would read

$$I_i^{(\mathcal{R};as)} = \frac{\left(\sum_x N_{x,i}^*\right)^2}{\sum_x N_{x,i}^{*\,2} \frac{1}{I_{x,i}}}, \quad \text{(D8)}$$

('$as$' stands for *asynchronous space*.) On the contrary, if the local dynamics were perfectly synchronous ($\text{corr}_t\left(N_x, N_y\right) = 1$), the regional invariability would be

$$I_i^{(\mathcal{R};ss)} = \frac{\left(\sum_x N_{x,i}^*\right)^2}{\sum_x \sum_y N_{x,i}^* N_{y,i}^* \frac{1}{\sqrt{I_{x,i}}} \frac{1}{\sqrt{I_{y,i}}}} = \frac{\left(\sum_x N_{x,i}^*\right)^2}{\left(\sum_x N_{x,i}^* \frac{1}{\sqrt{I_{x,i}}}\right)^2}, \quad \text{(D9)}$$

('$ss$' stands for *synchronous space*).

We define the typical correlation between different locations for species $i$ as

$$\bar{c}_i \equiv \frac{\sum_x \sum_{y \neq x} N_{x,i}^* N_{y,i}^* \text{corr}_t\left(N_{x,i}, N_{y,i}\right) \frac{1}{\sqrt{I_{x,i}}} \frac{1}{\sqrt{I_{y,i}}}}{\sum_x \sum_{y \neq x} N_{x,i}^* N_{y,i}^* \frac{1}{\sqrt{I_{x,i}}} \frac{1}{\sqrt{I_{y,i}}}}, \quad \text{(D10)}$$

which is the spatial mean of the temporal correlation between different locations, with weights given by the local biomasses and invariabilities. Moving the denominator to the left side of the equation, Eq. (D10) would read

$$\bar{c}_i \left(\sum_x \sum_{y \neq x} N_{x,i}^* N_{y,i}^* \frac{1}{\sqrt{I_{x,i}}} \frac{1}{\sqrt{I_{y,i}}}\right) = \sum_x \sum_{y \neq x} N_{x,i}^* N_{y,i}^* \text{corr}_t\left(N_{x,i}, N_{y,i}\right) \frac{1}{\sqrt{I_{x,i}}} \frac{1}{\sqrt{I_{y,i}}}. \quad \text{(D11)}$$

Now, if we express the sums $\sum_x \sum_{y \neq x} A_{xy}$ as $\sum_x \sum_y A_{x,y} - \sum_x A_{x,x}$, this expression would read

$$\bar{c}_i \left( \sum_x \sum_y N_{x,i}^* \, N_{y,i}^* \frac{1}{\sqrt{I_{x,i}}} \frac{1}{\sqrt{I_{y,i}}} - \sum_x {N_{x,i}^*}^2 \frac{1}{I_{x,i}} \right) = \sum_x \sum_y N_{x,i}^* \, N_{y,i}^* \, \mathrm{corr}_t \left( N_{x,i}, N_{y,i} \right) \frac{1}{\sqrt{I_{x,i}}} \frac{1}{\sqrt{I_{y,i}}}$$

$$- \sum_x {N_{x,i}^*}^2 \frac{1}{I_{x,i}}. \tag{D12}$$

Moreover, we also have that

- $\sum_x {N_{x,i}^*}^2 \frac{1}{I_{x,i}} = \frac{\left( \sum_x N_{x,i}^* \right)^2}{I_i^{(\mathcal{R};as)}}$ (Eq. (D8)),

- $\sum_x \sum_y N_{x,i}^* \, N_{y,i}^* \frac{1}{\sqrt{I_{x,i}}} \frac{1}{\sqrt{I_{y,i}}} = \frac{\left( \sum_x N_{x,i}^* \right)^2}{I_i^{(\mathcal{R};ss)}}$ (Eq. (D9))

- $\sum_x \sum_y N_{x,i}^* \, N_{y,i}^* \, \mathrm{corr}_t \left( N_{x,i}, N_{y,i} \right) \frac{1}{\sqrt{I_{x,i}}} \frac{1}{\sqrt{I_{y,i}}} = \frac{\left( \sum_x N_{x,i}^* \right)^2}{I_i^{(\mathcal{R})}}$ (Eq. (D7))

Hence, Eq. (D12) can be written as

$$\bar{c}_i \left( \frac{\left( \sum_x N_{x,i}^* \right)^2}{I_i^{(\mathcal{R};ss)}} - \frac{\left( \sum_x N_{x,i}^* \right)^2}{I_i^{(\mathcal{R};as)}} \right) = \frac{\left( \sum_x N_{x,i}^* \right)^2}{I_i^{(\mathcal{R})}} - \frac{\left( \sum_x N_{x,i}^* \right)^2}{I_i^{(\mathcal{R};as)}}$$

$$I_i^{(\mathcal{R})} = \left[ (1 - \bar{c}_i) \frac{1}{I_i^{(\mathcal{R};as)}} + \bar{c}_i \frac{1}{I_i^{(\mathcal{R};ss)}} \right]^{-1}. \tag{D13}$$

*I.e.*, the regional invariability can be computed as a harmonic mean of the invariabilities for the asynchronous and synchronous cases, weighted by the typical spatial correlation.

Analogous computations would let as to express the community invariability as

$$I_x^{(\mathcal{C})} = \frac{\left( \sum_i N_{x,i}^* \right)^2}{\sum_i \sum_j N_{x,i}^* \, N_{x,j}^* \, \mathrm{corr}_t \left( N_{x,i}, N_{x,j} \right) \frac{1}{\sqrt{I_{x,i}}} \frac{1}{\sqrt{I_{x,j}}}} = \left[ (1 - \widetilde{c}_x) \frac{1}{I_x^{(\mathcal{C};ac)}} + \widetilde{c} \frac{1}{I_x^{(\mathcal{C};sc)}} \right]^{-1}, \tag{D14}$$

with $\widetilde{c}$ the typical correlation between different species, and '$ac$' and '$sc$' standing for *asynchronous community* and *synchronous community*, respectively.

Eq. (D9) shows us that for perfectly synchronous local dynamics, the regional species invariability is the square of the harmonic mean of the square root of the local species invariabilities, weighted by the equilibrium local biomass densities. Conversely, if the local species biomass dynamics is spatially asynchronous (Eq. (D8)), the regional species invariability of the whole spatial network is

$$I_i^{(\mathcal{R};as)} = \frac{\left( \sum_x N_{x,i}^* \right)^2}{\sum_x \left( N_{x,i}^* \right)^2 \frac{1}{I_{x,i}}} = n_L \frac{1}{1 + \left( \frac{\sigma_{N_i^*}}{\mu_{N_i^*}} \right)^2} \frac{\sum_x \left( N_{x,i}^* \right)^2}{\sum_x \left( N_{x,i}^* \right)^2 \frac{1}{I_{x,i}}}, \tag{D15}$$

For the asynchronous-space case, the regional invariability is proportional to the number of locations $n_L$ (Fig. 6 of the main text, Fig. S5), and to the harmonic mean of local species invariabilities weighted by the squared local species biomasses; and it is modulated by the spatial variance $\sigma^2_{N_i^*}$ and mean $\mu_{N_i^*}$ of the average local biomasses of species $i$. Invariability for asynchronous spaces is always greater than the invariability for the synchronous space cases. Indeed, the synchronous (Eq. (D9)) and asynchronous (Eq. (D8)) cases have the same numerator $\left(\left(\sum_x N_{x,i}^*\right)^2\right)$, while the denominator for the synchronous case $\left(\left(\sum_x N_{x,i}^* \frac{1}{\sqrt{I_{x,i}}}\right)^2 = \sum_x \left(N_{x,i}^*\right)^2 \frac{1}{I_{x,i}} + \sum_x \sum_{y \neq x} N_{x,i}^* N_{y,i}^* \frac{1}{\sqrt{I_{x,i} I_{y,i}}}\right)$ is larger than for the asynchronous case $\left(\sum_x \left(N_{x,i}^*\right)^2 \frac{1}{I_{x,i}}\right)$.

## Appendix E: Model simulations

To test the generality of the results discussed in the manuscript, we performed numerical simulations for the community dynamics of 10 competitor species in random spatial networks. For such numerical simulations, we consider three contributions to the community dynamics: the deterministic local dynamics, the deterministic dispersal dynamics, and a term with the environmental stochasticity,

$$\frac{dN_{x,i}(t)}{dt} = \left.\frac{dN_{x,i}(t)}{dt}\right|_{local} + \left.\frac{dN_{x,i}(t)}{dt}\right|_{disp} + \left.\frac{dN_{x,i}(t)}{dt}\right|_{env}. \tag{E1}$$

For the local deterministic dynamics, we assume the Lotka-Volterra competition model

$$\left.\frac{dN_{x,i}(t)}{dt}\right|_{local} = N_{x,i}(t)\left(r_{x,i} - \sum_j \alpha_{x,ij}\, N_{x,j}(t)\right). \tag{E2}$$

The local per-capita growth rates of each of the competitor at each of the locations were sampled from a normal distribution with mean 1 and standard deviation of $1/10$, $r_{x,i} \sim \mathcal{N}\left(\mu = 1, \sigma = \frac{1}{10}\right)$. The diagonal terms of the interaction strengths were randomly sampled from a normal distribution with mean $1/100$ and a standard deviation of $\frac{1}{10}$ of such mean, $\alpha_{x,ii} \sim \frac{1}{100}\mathcal{N}\left(\mu = 1, \sigma = \frac{1}{10}\right)$. And the off-diagonal terms were samples from a normal distribution with smaller means and variances, $\alpha_{x,ij} \sim \frac{1}{1000}\mathcal{N}\left(\mu = 1, \sigma = \frac{1}{10}\right)$. Any negative value of any of the local growth rates or of the interaction strengths was converted to $\frac{1}{100}$ of the mean of the distribution.

For the dispersal, we generate random networks with different number of nodes, and connections between some of the nodes (Fig. S1). Individuals of each of the population will be able to diffuse through connected nodes of the spatial network, with the expression

$$\left.\frac{dN_{x,i}(t)}{dt}\right|_{disp} = -\sum_y m_{x\to y,i}\, N_{x,i}(t) + \sum_z m_{z\to x,i}\, N_{z,i}(t) \tag{E3}$$

where the first term represents the emigration from the local node $x$ to each of the connected nodes, and the second term represents immigration to $x$ from any connected node. The elements $m_{x \to y, i}$ are equal to zero for any pair of nodes not connected by an edge; while for connected nodes they are sampled from a normal distribution $m_{x \to y, i} \sim \frac{1}{d_{xy}} \mathcal{N}\left(\mu = 1, \sigma = \frac{1}{10}\right)$, being . $d_{xy}$ the spatial distance between the nodes.

Finally, for the environmental stochasticity term we have employed

$$\left.\frac{dN_{x,i}(t)}{dt}\right|_{env} = N_{x,i}(t)\, \sigma_{x,i}\, \frac{dB_{x,i}}{dt}(t) \tag{E4}$$

where the amplitudes of the environmental stochasticity $\sigma_{x,i}$ have been sampled from an exponential distribution with mean $\frac{1}{20}$. The terms $\frac{dB_{x,i}}{dt}(t)$ can be taken as independent white noises for each population and location, for studying the dynamics in the spatial and population asynchronous case; or as a unique common white noise shared by all populations and locations, for studying the dynamics in a perfectly synchronous scenario.

For estimating the invariability of spatial community networks, we run the community dynamics for a time span of 300 units, with time-steps of 0.25, from initial populations equal to the species carrying capacities (in absence of interspecific competition). Then, we select the last half of the time series, in order to remove the transient dynamics. From these truncated time series, we compute the local population temporal means and variances. All the numerical simulation were done in Python 3.7 (Python Core Team, 2019).

For estimating the growth rate, initial resilience, and resistance of spatial community networks, starting again at populations equal to the local carrying capacities, we run the simulations for 100 time units (with time-steps of 0.25) to reach the network equilibrium state. Then, from time 100 we reduce each of the relative local population growth rates a quantity sampled from a normal distribution, $r_{x,i} \to r_{x,i}\left(1 - \xi_{x,i}\right)$, with $\xi_{x,i} \sim \frac{1}{2}\mathcal{N}\left(\mu = 1,\ \sigma = \frac{1}{4}\right)$, and we run the community dynamics from time 100 to time 200 (to let the network reach again its new equilibrium state), again with time steps equal to 0.25. Resistance is estimated from the relative decrease in the population abundances at times 100 (equilibrium before the perturbation) and times 200 (equilibrium after the perturbation). Finally, at time 200 we set again the local population growth rates to their pre-perturbated values, and we run the community dynamics until time 300 (to check that we return the pre-perturbed equilibrium state). From these last time series, growth rate is estimated as the initial return rate to the pre-perturbed equilibrium in the first time-step.

## Appendix F: Scaling of not-normalized invariance

We define the invariance as the inverse of the temporal variance of the population,

$$\mathcal{I} = \frac{1}{\mathrm{var}_t\big(N(t)\big)}\,. \tag{F1}$$

This invariance can be considered as another component of the ecosystem stability. It is analogous to the invariability (Eq. (5) of the main text), but without normalizing the variance of the abundances with the temporal mean of the biomasses.

If we want to study the invariance of a (either spatial or ecological) network, we define it as the inverse of the variance in the network total biomass,

$$\mathcal{I}_G = \frac{1}{\mathrm{var}_t(\sum_a N_a(t))} \, , \tag{F2}$$

where $a$ holds for the node of the network, which can represent either a species or a location, and $\mathcal{I}_G$ holds for the global (either community or regional) invariance. Eq. (F2) can be rewritten as

$$\mathcal{I}_G = \frac{1}{\sum_a \mathrm{var}_t(N_a(t)) + \sum_a \sum_{b \neq a} \mathrm{cov}_t(N_a(t), N_b(t))} \, , \tag{F3}$$

with $\mathrm{cov}_t(N_a(t), N_b(t))$ the temporal covariance between the nodes $a$ and $b$.

Now, we distinguish two scenarios

- In the case of asynchronous dynamics between each pair of different nodes of the network, $\mathrm{cov}_t(N_a(t), N_b(t)) = 0$ for $a \neq b$. Hence, since the average local variance is simply $\overline{\mathrm{var}_t(N_a(t))} = \sum_a \mathrm{var}_t(N_a(t))/n$ (with $n$ the number of the nodes of the network), the global invariance would simply be

$$\mathcal{I}_G^A = \frac{1}{n \; \overline{\mathrm{var}_t(N_a(t))}} = \frac{1}{n} \frac{1}{(1/\mathcal{I}_a)} = \frac{\langle \mathcal{I}_a \rangle}{n} \, , \tag{F4}$$

  where $\langle \mathcal{I}_a \rangle$ is the harmonic mean of the invariances at the nodes of the network.

- In the case of perfect synchronous dynamics, $\overline{\mathrm{cov}_t(N_a(t), N_b(t))} = \overline{\mathrm{var}_t(N_a(t))}$ for each pair of nodes. Hence, the global invariance would be

$$\mathcal{I}_G^S = \frac{1}{\left(n + n(n-1)\right) \overline{\mathrm{var}_t(N_a(t))}} = \frac{1}{n^2} \frac{1}{(1/\mathcal{I}_a)} = \frac{\langle \mathcal{I}_a \rangle}{n^2} \, . \tag{F5}$$

In both synchronous and asynchronous regimes, the global invariance decreases with the number of nodes. Hence, networks are fundamentally less invariant (present higher variances) than their nodes.

## Appendix G: Scaling of not-normalized resistance

We define the not-normalized resistance as inversely proportional to the change of the biomass caused by a perturbation at time $t_0$,

$$\omega = \frac{1}{N(t_0) - N(t_0 + \delta t)} \, . \tag{G1}$$

This not-normalized resistance is equal to the normalized resistance that we employed throughout the main text (Eq. (2) of the main text), without normalizing it with the biomass before the perturbation.

If we define the local not-normalized resistance as the inverse of the biomass change at a node $a$ caused by a perturbation, we can simply define the global not-normalized resistance of the network as the not-normalized resistance of the total biomass

$$\omega_G = \frac{1}{\sum_a N_a(t_0) - \sum_a N_a(t_0 + \delta t)} \, , \tag{G2}$$

which can be rewritten as

$$\omega_G = \frac{1}{\sum_a \left(N_a(t_0) - N_a(t_0 + \delta t)\right)} = \frac{1}{\sum_a \frac{1}{\omega_a}} = \frac{\langle \omega_a \rangle}{n} \, . \tag{G3}$$

That is, the global estimate of the not-normalized resistance is equal to the harmonic mean of the local estimates, divided by the number of nodes of the network. Thus, networks would have fundamentally smaller values of this estimate than the nodes forming these networks, so it is not a scale-free property.

**Supplementary Figures**

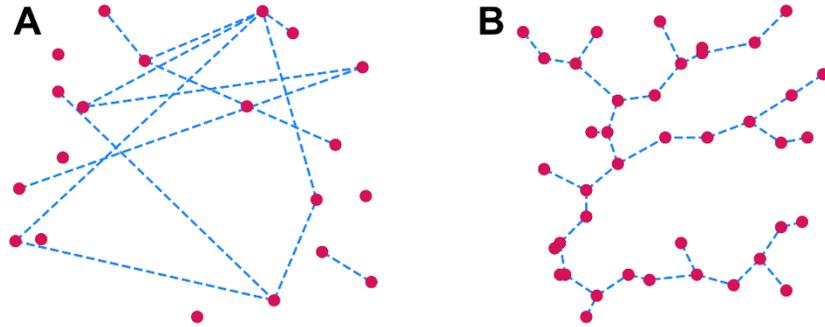

**Figure S1**. Spatial random networks employed for the computation of regional estimates of the different stability estimates. (**A**) Random spatial networks. The nodes and connections between the nodes are constructed as Erdős-Rényi random graphs $G(N, p)$, with $N$ the number of edges and $p$ the probability of presence of each possible edge between the nodes (taken as $p = \frac{1}{10}$ for all the plots depicted through the text). These graphs were generated using the Networkx package (Hagberg et al., 2008) in Python 3.7 (Python Core Team, 2019). The positions of the nodes (both x and y axes) are assigned from a uniform distribution $U(0, N)$, to ensure that bigger random spatial networks cover larger regions. In each of the patches, local dynamics is governed by a Lotka-Volterra Model. Moreover, diffusion is implemented at each time step between each pair of connected nodes. (**B**) Random spatial dendritic networks, generated in R (R Core Team, 2020) with the OCNet package (Carraro et al., 2020). These dendritic networks resemble those of riverine structures, and present a higher spatial order than the previous random networks. As before, local dynamics is assumed to be Lotka-Volterra, and diffusion is possible just between connected nodes of the network.

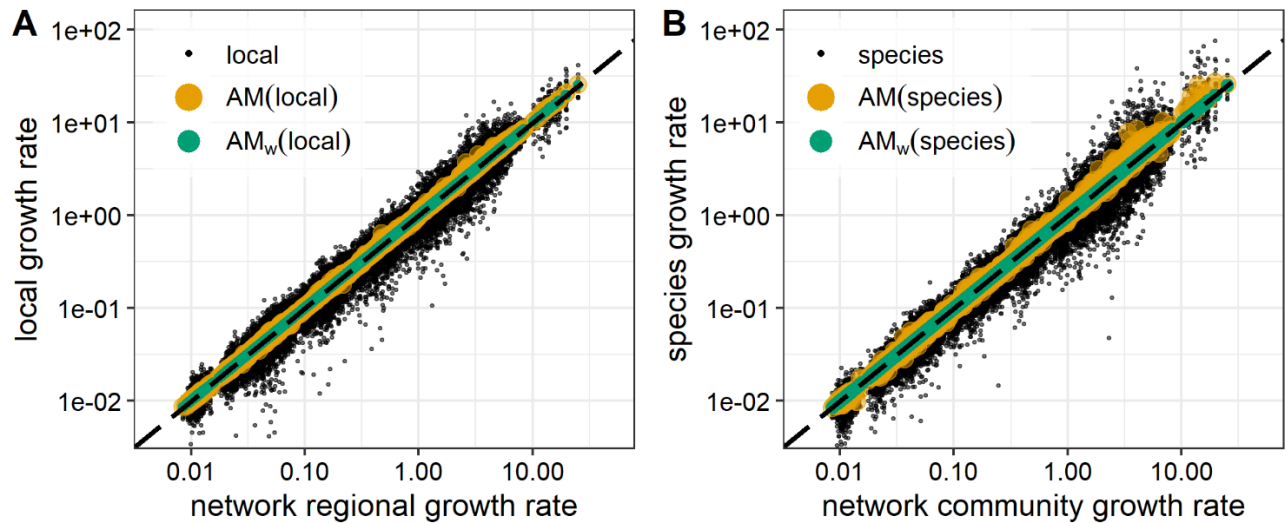

**Figure S2:** Analogous to Fig. 2 of the main text, but for random communities of 10 competitors in 10 node random dendritic networks (Fig. S1B). Regional (**A**) and community (**B**) growth rates, compared to local and species growth rates and to their unweighted arithmetic mean (AM) (yellow circles), and biomass weighted arithmetic mean ($AM_w$) (green circles). AM and $AM_w$ values closer to the identity line (black, dashed) estimate more precisely the regional (panel A) and community growth rate (panel B). Black dots are the growth rates of individual localities (A) or species (B) on which the means AM and $AM_w$ were computed. $AM_w$ are found to be better estimators of the growth rate, as expected from the results shown in the text

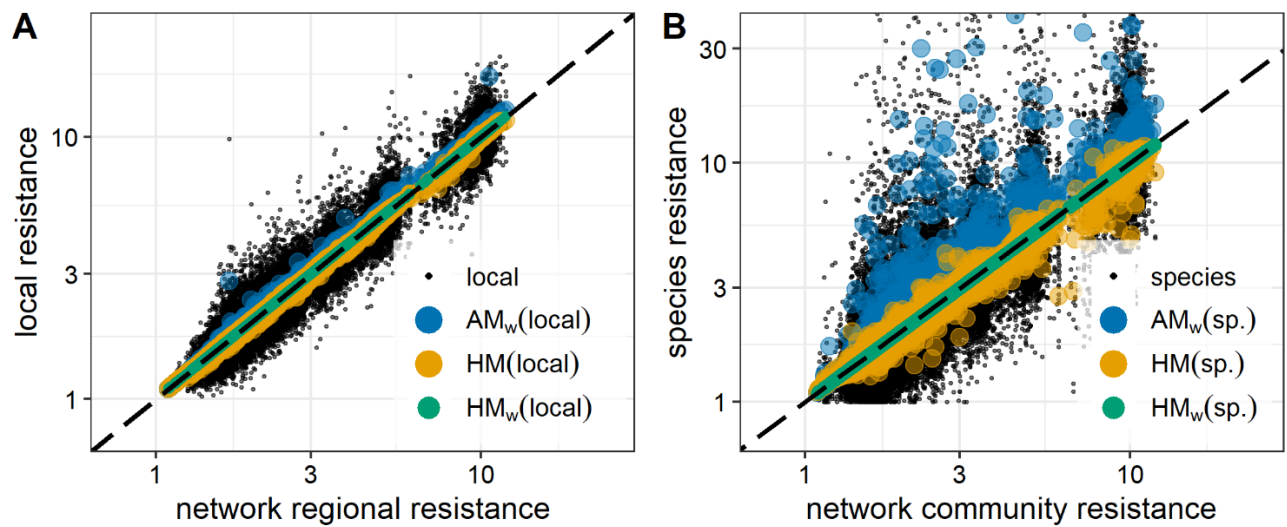

**Figure S3:** Analogous to Fig. 4 of the main text, but for random communities of 10 competitors in 10 node random dendritic networks (Fig. S1B). Regional (**A**) and community (**B**) resistances, compared to local and species resistances and to their biomass weighted arithmetic mean ($AM_w$) (blue circles), unweighted harmonic mean (HM) (yellow circles), and biomass weighted harmonic mean ($HM_w$) (green circles). $AM_w$, HM, and $HM_w$ values closer to the identity line (black, dashed) estimate more precisely the regional (panel A) and community resistance (panel B). Black dots are the resistances of the individual localities (A) or species (B) on which the means were computed. $HM_w$ are found to be better estimators for the resistance, as expected from the results shown in the text.

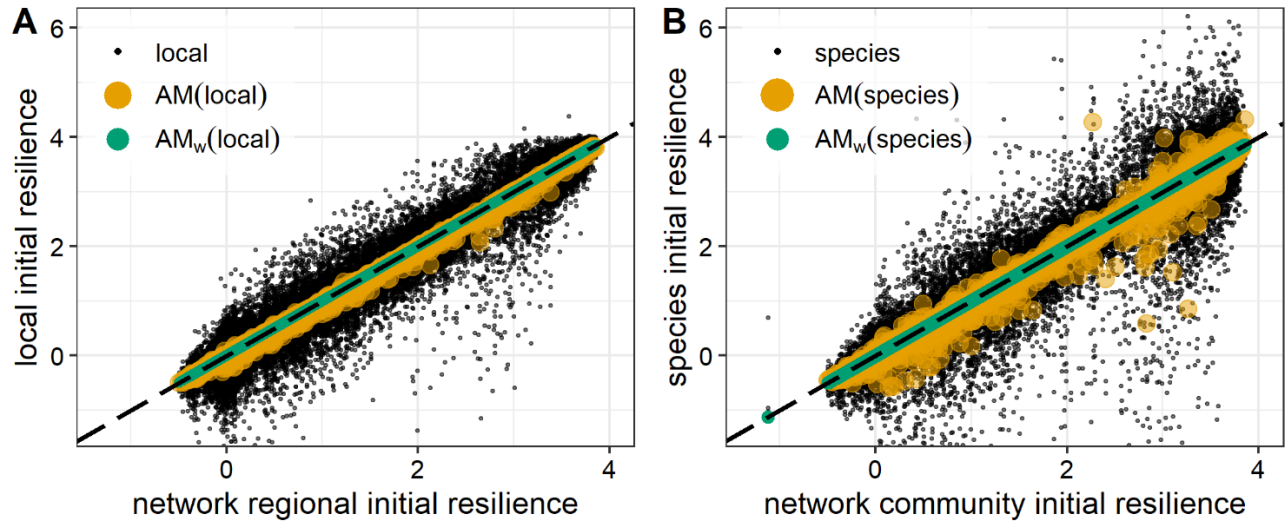

**Figure S4:** Analogous to Fig. 5 of the main text, but for random communities of 10 competitors in 10 node random dendritic networks (Fig. S1B). Regional **(A)** and community **(B)** initial resiliences, compared to local and species initial resiliences and to their unweighted arithmetic mean (AM) (yellow circles) and biomass weighted arithmetic mean ($AM_w$) (green circles). AM and $AM_w$ values closer to the identity line (black, dashed) estimate more precisely the regional (panel A) and community initial resilience (panel B). Black dots are the initial resiliences of the individual localities (A) or species (B) on which the means were computed. $AM_w$ are found to be better estimators of the initial resilience, as expected from the results shown in the text.

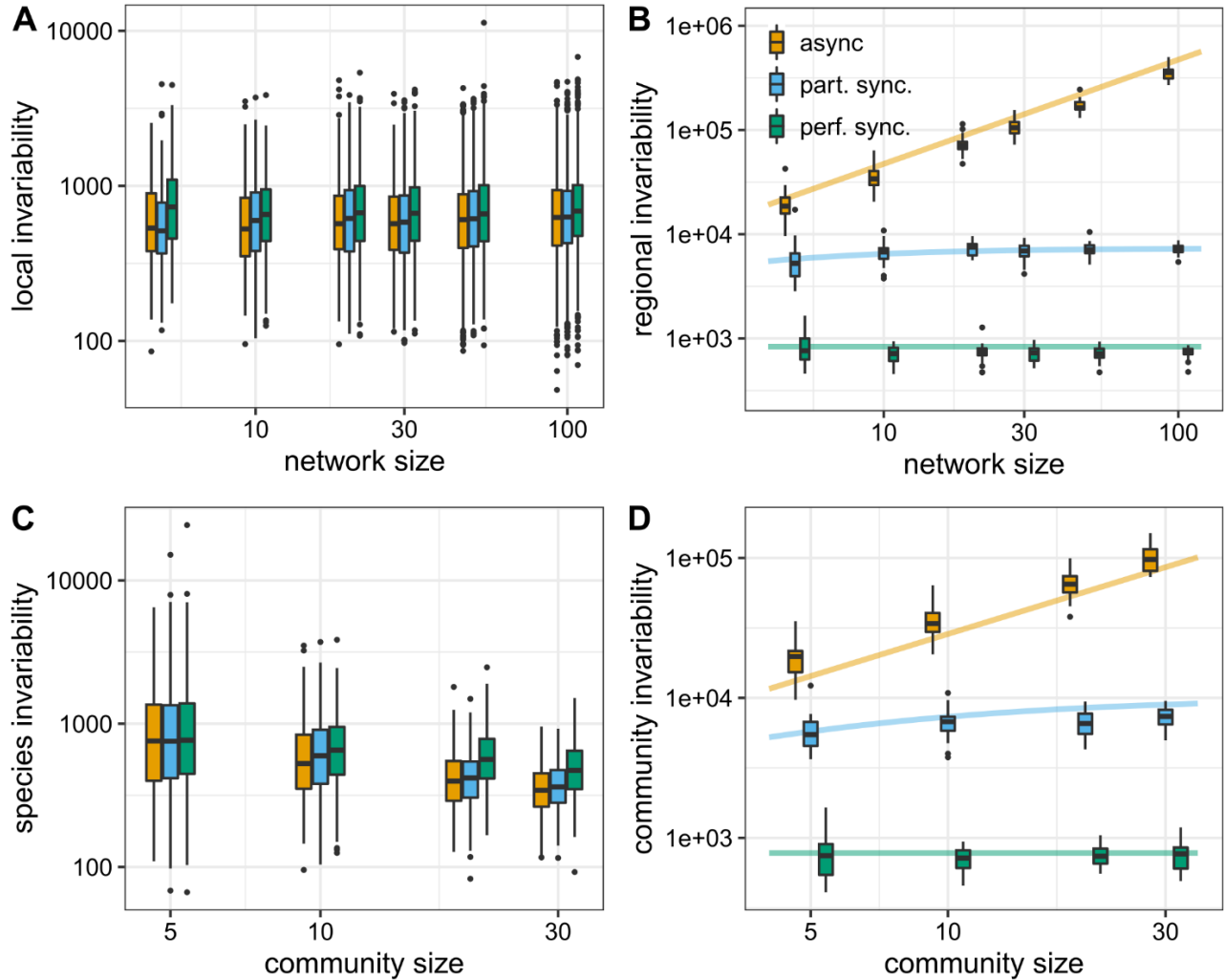

**Figure S5**: Analogous to Fig. 6 of the main text, but for random spatial dendritic networks (Fig. S1B). Local (**A**) and regional (**B**) invariability estimates in random dendritic networks of random communities of 10 competitor species, for different sizes of the spatial network; and species (**C**) and community (**D**) invariability estimates of random communities of competitors at 10-node random dendritic networks, for different number of species forming the communities. In panels A and B, we have considered three different scenarios: asynchronous local dynamics ($\bar{c} = 0$, yellow box plots), partially-synchronous local dynamics ($\bar{c} = 0.44$, blue), and perfectly-synchronous local dynamics ($\bar{c} = 1$, green). In panels C and D, we have considered the cases of asynchronous ($\tilde{c} = 0$), partially-synchronous ($\tilde{c} = 0.17$) and perfectly-synchronous ($\tilde{c} = 1$) species dynamics. Solid lines depict the invariabilities predicted by analytical expressions (Eqs. (20)-(22), and analogous expressions for community invariability)

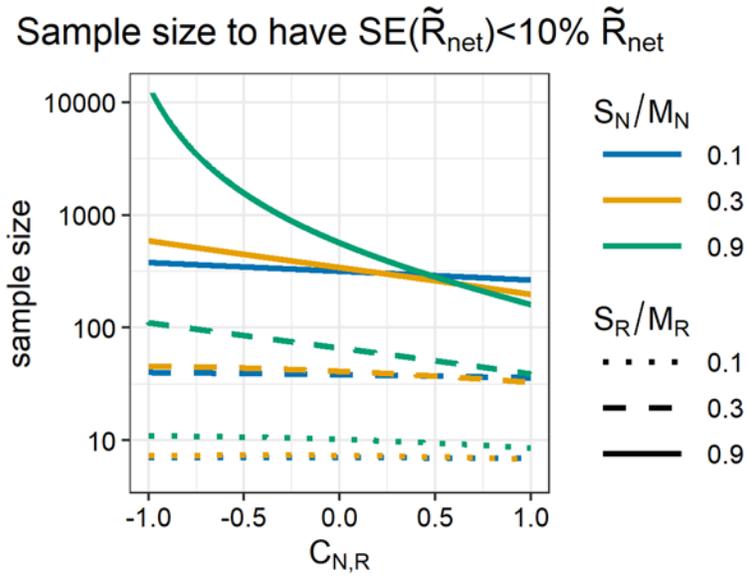

**Figure S6**: Dependence of the sample size required to control the relative standard error of the network growth rate with the correlation of the nodes' biomasses and growth rates, $C_{N,R}$, for different coefficients of variation of growth rate ($S_R/M_R$) and biomass ($S_N/M_N$). Confirming results of Fig. 3 of the main text, the required sample size increases mainly with the coefficient of variation of the node growth rates, and to a lesser extent with the coefficient of variation of the node biomasses except for the case of high anticorrelation. Moreover, the required sample size decreases with the correlation of biomasses and growth rates.

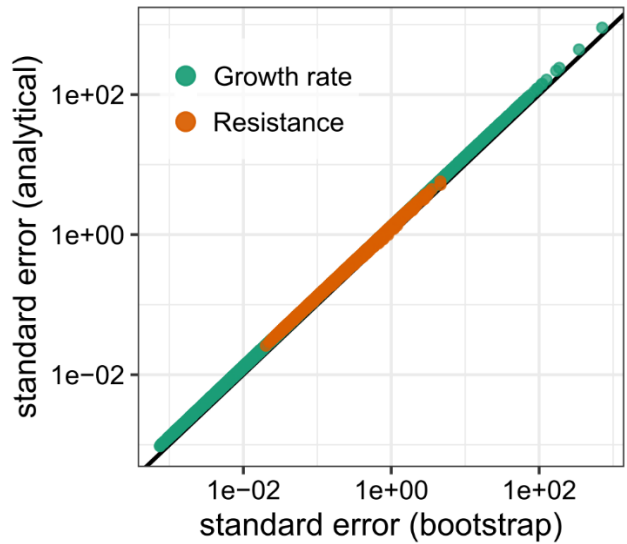

**Figure S7**: The analytical approximation for the standard error of regional growth rate (Eq. (11) of the main text) and resistance (Eq. (A5) of Appendix A) compares well to the numerically simulated standard errors obtained with bootstrapping techniques (Efron and Tibshirani, 1985; Hesterberg, 2011). These results have been obtained for the population dynamics of a unique species in random spatial networks of 20 nodes (Fig. S1A).

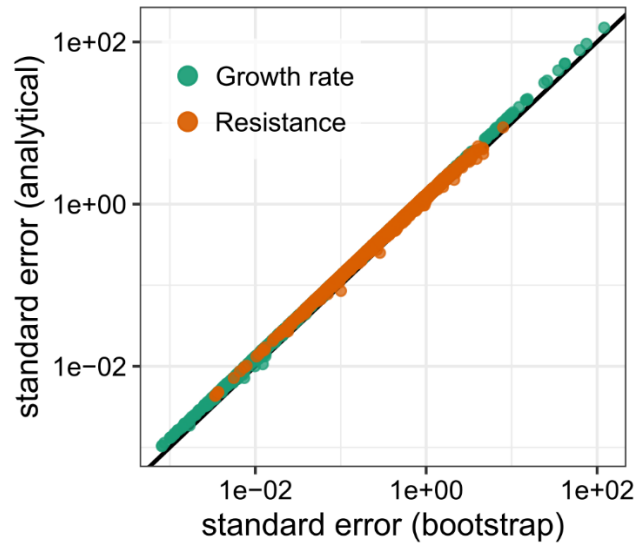

**Figure S8:** Analogous to Fig. S7, but for random spatial dendritic networks (Fig. S1B). The analytical expressions compare well to the numerically computed standard errors.

## Bibliography

Bohrnstedt, G. W., and Goldberger, A. S. (1969). On the Exact Covariance of Products of Random Variables. *J. Am. Stat. Assoc.* 64, 1439. doi:10.2307/2286081.

Carraro, L., Bertuzzo, E., Fronhofer, E. A., Furrer, R., Gounand, I., Rinaldo, A., et al. (2020). Generation and application of river network analogues for use in ecology and evolution. *Ecol. Evol.* 10, 7537–7550. doi:10.1002/ece3.6479.

Cochran, W. G. (1977). *Sampling Techniques*. 3rd editio. New York: John Wiley & Sons.

Efron, B., and Tibshirani, R. (1985). The Bootstrap Method for Assessing Statistical Accuracy. *Behaviormetrika* 12, 1–35. doi:10.2333/bhmk.12.17_1.

Gatz, D. F., and Smith, L. (1995). The standard error of a weighted mean concentration-I. Bootstrapping vs other methods. *Atmos. Environ.* 29, 1185–1193. doi:10.1016/1352-2310(94)00210-C.

Hagberg, A. A., Schult, D. A., and Swart, P. J. (2008). Exploring network structure, dynamics, and function using NetworkX. *7th Python Sci. Conf. (SciPy 2008)*, 11–15.

Hesterberg, T. (2011). Bootstrap. *Wiley Interdiscip. Rev. Comput. Stat.* 3, 497–526. doi:10.1002/wics.182.

Norris, N. (1940). The Standard Errors of the Geometric and Harmonic Means and Their Application to Index Numbers. *Ann. Math. Stat.* 11, 445–448. doi:10.1214/aoms/1177731830.

Python Core Team (2019). Python: A dynamic, open source programming language. Available at: https://www.python.org/.

R Core Team (2020). R: A Language and Environment for Statistical Computing. Available at: https://www.r-project.org/.